\documentclass[preprint,11pt,usletter]{elsarticle}

\usepackage{lineno,hyperref}
\usepackage{float}
\usepackage{fullpage}
\usepackage{multirow}
\usepackage{amsmath}
\usepackage{fontenc}
\usepackage{subfig}
\usepackage{bm}
\usepackage{amssymb}
\usepackage{graphicx}
\usepackage{multirow}
\usepackage{mathtools}
\DeclarePairedDelimiter{\norm}{\lVert}{\rVert}
\usepackage{graphicx}
\usepackage{stackengine}
\usepackage{wrapfig}
\usepackage{amssymb}
\usepackage[dvipsnames]{xcolor}
\usepackage{algorithm2e}
\DontPrintSemicolon
\SetNoFillComment
\usepackage{xcolor, soul}
\usepackage{array}
\newcolumntype{L}{>{\centering\arraybackslash}m{4.5cm}}
\newcolumntype{B}{>{\centering\arraybackslash}m{3.5cm}}
\newcolumntype{R}{>{\centering\arraybackslash}m{2.5cm}}
\newcolumntype{S}{>{\centering\arraybackslash}m{1.6cm}}

\def\bfm#1{{\bf #1}}

\def\bfs#1{\mbox{\boldmath{$ #1 $}}}

\newcommand\rd{d}

\definecolor{fgwhite}{rgb}{1,1,1}     
\definecolor{fgred}{rgb}{0.8,0,0}     
\definecolor{Henna}{rgb}{0.63,0.63,0.08}     
\definecolor{fggreen}{rgb}{0,0.5,0}     
\definecolor{fgpurple}{rgb}{0.5,0,1}     
\definecolor{fggray}{rgb}{0.6,0.6,0.7}     
\definecolor{bggreen}{rgb}{0.8,1,0.8}     
\definecolor{fgblue}{rgb}{0,0,0.7}     
\definecolor{bgblue}{rgb}{0.9,0.9,1}     
\definecolor{fgclay}{rgb}{0.51,0.25,0.04}     
\definecolor{paleyellow}{rgb}{0.9,0.9,0.06}  

\sethlcolor{green}

\newcommand{\rev}[1]{#1}

\newcommand{\GP}{\mathcal{GP}}
\newcommand{\N}{\mathcal{N}}

\setcitestyle{square,comma,numbers,sort&compress}
\usepackage{bbm}
\usepackage{bbold}

\journal{-}

\biboptions{numbers,sort&compress}

\begin{document}

\begin{frontmatter}
\title{Gaussian process regression + deep neural network autoencoder\\for probabilistic surrogate modeling in nonlinear mechanics of solids}

\author[mymainaddress]{Saurabh Deshpande}

\author[Exeter]{Hussein Rappel}

\author[Mark]{Mark Hobbs}

\author[mymainaddress]{St\'ephane P.A. Bordas}

\author[mymainaddress,mysecondaryaddress,mythirdaddress]{Jakub Lengiewicz}

\cortext[mycorrespondingauthor]{Corresponding author}
\ead{jleng@ippt.pan.pl}

\address[mymainaddress]{Department of Engineering; Faculty of Science, Technology and Medicine; University of Luxembourg}
\address[Exeter]{Department of Engineering, Faculty of Environment, Science and Economy, University of Exeter, UK}
\address[Mark]{Future Methods, Rolls-Royce plc, Derby, UK}
\address[mysecondaryaddress]{Institute of Fundamental Technological Research, Polish Academy of Sciences, Poland}
\address[mythirdaddress]{Luxembourg Institute of Science and Technology, Esch-sur-Alzette, Luxembourg}


\begin{abstract} 

Many real-world applications demand accurate and fast predictions, as well as reliable uncertainty estimates. However, quantifying uncertainty on high-dimensional predictions is still a severely under-investigated problem, especially when input-output relationships are non-linear. To handle this problem, the present work introduces an innovative approach that combines autoencoder deep neural networks with the probabilistic regression capabilities of Gaussian processes. The autoencoder provides a low-dimensional representation of the solution space, while the Gaussian process is a Bayesian method that provides a probabilistic mapping between the low-dimensional inputs and outputs. We validate the proposed framework for its application to surrogate modeling of non-linear finite element simulations. Our findings highlight that the proposed framework is computationally efficient as well as accurate in predicting non-linear deformations of solid bodies subjected to external forces, all the while providing insightful uncertainty assessments.

\end{abstract}

\end{frontmatter}


\section{Introduction} 

In computational mechanics, high-fidelity modeling methods such as the finite element method (FEM) are standard for achieving detailed insights into the mechanical behaviors of materials and structures. Despite their capability to provide accurate predictions for complex non-linear problems, these methods are often hindered by the substantial computational effort they demand, thereby constraining their applicability in domains requiring real-time solutions. As a remedy, various approximate methods, such as surrogate modeling, have emerged as an indispensable tool to alleviate the computational burden associated with traditional simulations. By providing the balance between computational efficiency and the accuracy, these methods play a crucial role in scenarios requiring fast- or real-time responses.

In conventional methods like FEM, numerical challenges primarily arise from two main factors: the non-linearities present in the problem and the high number of unknowns (degrees of freedom) required to attain the desired level of prediction accuracy. The majority of approximate methods concentrate on addressing the latter aspect by directly reducing the number of unknowns of the problem. For instance, the conventional reduced ordered modelling (ROM) methods accomplish this by linearly projecting higher-order information to lower-dimensional space using dimensionality reduction techniques such as Principal Component Analysis (PCA) \citep{PCAFEM,AVERSANO2019422}, and Proper Orthogonal Decomposition (POD) \citep{PODSoft,PAPAVASILEIOU2023103938}. However, these approaches yield undesirable results when simulating highly nonlinear phenomena \citep{scholkopf1998nonlinear, POD_limitations, MOR2}. In response to the linearity limitation, numerous nonlinear dimensionality reduction techniques have been proposed~\cite{Mendez_2023,kernel_POD}. However, obtaining real-time simulations with these methods, in particular when considering high degree of non-linearity, remains a challenge~\citep{cotin}.

Rapid advancements in machine learning has naturally fueled developments to approximate modelling. There has been a widespread adoption of machine learning techniques to accelerate computationally intensive numerical methods \citep{computation8010015,annurev,GLUMAC2023110135, thesis-saurabh,doi:10.1126/science.adi2336}. In particular, deep learning (DL) models have been successfully used as accurate and computationally inexpensive surrogates to solve non-linear problems in mechanics \citep{Deshpande2023, vasilis, vasilisgraph, DESHPANDE2024108055, Frydrych2024}. In context of ROM, DL methods are being extensively used to find \emph{non-linear} reduced order representations of the \rev{high dimensional data \citep{wang2016auto, ShiZhouWang2024}.} At the same time, they are computationally inexpensive at the prediction stage, hence, they are highly suitable in a wide area of applications requiring real time simulations \citep{reddy2020analysis,fresca2021comprehensive}.

This work focuses on another important aspect, which is the capabilities of surrogate models to quantify uncertainty. Providing reliable uncertainty estimates alongside predictions is needed in various real-life applications, such as surgical simulations \citep{MAZIER2021110645} or autonomous driving \citep{9001195}, to reduce risks of harmful consequences in these crucial tasks. There exist many machine learning approaches allowing to quantify model- and data uncertainties. In deep learning, the most common approach is to use stochastic deep neural network architectures with relevant suitable training procedures. These are, for instance, \citep{bayesbackprop, dropout_bayes,kingma2022autoencodingvariationalbayes}. They are known to be capable of providing uncertainty information, however, they can be computationally much more intensive as compared to their deterministic versions. Another class of approaches, employed specifically in this work, are probabilistic models that are based on Gaussian process regression~\citep{C_Rasmussen_2006,S_Roger_2016} (GP regression or GPR). 
The excellent predictive power and intrinsic capability of GPRs to quantify uncertainties make them a perfect candidate for the application of this contribution. Indeed, GPs have been employed also in mechanics in a wide range of studies both as surrogate models \cite{M_Kennedy_2001, M_J_Bayarri_2007, P_Arendt_2012, pmlr-v202-vadeboncoeur23a} and as an approach to model spatially varying parameter fields \cite{P_Koutsourelakis_2009,10.1115/1.4044894,https://doi.org/10.1002/nme.6974}. Ding et al.~\citep{DING2023115855} employed a combination of GPR and PCA to model displacement fields in problems with nonlinear materials (i.e., elastoplastic material), \citep{Jidling1073616} and \citep{POLONI2023110056} used GPR to assess uncertainties in the full-field measurement experimental context.

Despite being a powerful technique for constructing probabilistic surrogate models, GPR is known for its poor computational scaling with respect to number data points \citep{liu2020gaussian}. The scalability issue is substantially more significant for multioutput GPs (MOGP), where surrogate models gives multiple outputs simultaneously (i.e., $\mathcal{O}(N^{3}M^{3})$, where $N$ and $M$ denote the number of data points and the dimension of the multiple output, respectively~\citep{pmlr-v119-bruinsma20a}. Numerous approximated techniques have been developed to significantly reduce this very poor scaling. The primary effort was to reduce the scaling with respect to $N$, making the GPR more suitable for big data, see~\citep{liu2020gaussian} for a more extensive review. For the multi-output GPs, the most common simplification is to apply independent GPs for each dimension of the output (Independent Multi-Output Models), which reduces the scaling to $N^3M$. 

The framework proposed in this contribution focuses on alleviating part of the scaling issue by directly reducing the dimension of outputs,~$M$. This is achieved by applying the compressed latent representation of outputs via, so called, autoencoder deep neural networks---a strategy akin to approaches found in related works~\citep{MAULIK2021132797, DONNELLY2024107536}.
Autoencoder neural networks, as an unsupervised learning technique, have been commonly adopted to efficiently learning compressed representations across various domains \citep{masci2011stacked, liou2014autoencoder, doi:10.1177/1475921720934051}, also in the context of computational fluid dynamics \citep{pant2021deep,murata2020nonlinear} and computational solid mechanics \citep{fresca2022deep,SHINDE202394}. The primary advantage of autoencoder DNNs is their ability to reduce the dimensionality of data in a non-linear way~\citep{auto_hinton}. As such, they perform much better than POD or PCA when high non-linearities are present in data, \rev{see, e.g., the study~\citep{8001963}. Despite the known drawback of DNNs vs. PCA regarding their reduced interpretability of latent representations, in the present work we use DNNs} as an integral component of the proposed probabilistic framework. \rev{The reason for that is to provide a more general framework, capable of handling a wider class of datasets.}  

In the present work, we combine GPR with autoencoder DNNs to provide surrogate model for force-displacement predictions in mechanics of solids. In the proposed framework, we reduce the high-dimensional 2D/3D finite element solutions (full-field displacement) into their reduced representation by applying autoencoder neural networks, and then use GPR to provide probabilistic mapping between the input forces and the reduced-order displacements. During the prediction phase, for a given unseen force, GPR predicts the reduced-order (latent space) displacement which is then projected to the full field using the decoder part of autoencoder network. \rev{Such general idea to combine GPR with autoencoder DNNs is not new, and has been recently utilized in different applications and contexts. For instance, in the context of high-dimensional outputs, similar strategy has been applied to the spatiotemporal climate modeling~\citep{DONNELLY2024107536}, in which the full-field climate data is compressed using a DNN autoencoder while GP is used to forecast weather one step ahead. In the context of mechanics, in the works \citep{LI2020106399, LI2022115130}, techniques combining GPR and autoencoders have been developed for reliability analysis, and demonstrated for the problem of predicting structural compliance of 2D beams. In the present work, we extend the application to non-linear deformation of 3D solids with unstructured meshes. We emphasize on studying the contribution of errors by individual components (GP and autoencoder), and we provide interpretation of uncertainties in relation to the extrapolated region.} We demonstrate the robustness and versatility of our framework by applying it to synthetic datasets generated from the non-linear finite element simulations. Additionally, the source codes and datasets utilized in this work are provided in the following repository: \href{https://github.com/saurabhdeshpande93/gp-auto-regression}{https://github.com/saurabhdeshpande93/gp-auto-regression}.

The remainder of this paper is organized as follows. In Section~\ref{sec: methodology} we introduce the surrogate modeling framework, and describe its constituent parts. In Section~\ref{sec: results} we present two benchmark examples and study the performance of the framework. The conclusions and future research directions are outlined in Section~\ref{sec: conclusion}.


\section{Methodology}
\label{sec: methodology}

As outlined in the introduction, in this work, we focus on the problem of predicting non-linear deformation of solids under external loads. The goal is to introduce surrogate machine learning framework that is capable of accurate and fast predictions of full field displacements along with the associated uncertainties. The core idea is to combine the autoencoder deep neural network with the Gaussian Process Regression, see Section~\ref{sec: autoencoder model} and Section~\ref{sec: GP}, respectively. The training of both models is based on the numerically generated force-displacement dataset
\begin{equation}
    \mathcal{D}^f=\{(\bfm{f}_i,\bfm{u}_i)\}_{i=1,..,N},
\end{equation} 
which is obtained with the classical FEM, see Section~\ref{sec: fem_description}.

\begin{figure}[t]
     \centering
     \includegraphics[width=0.8\textwidth]{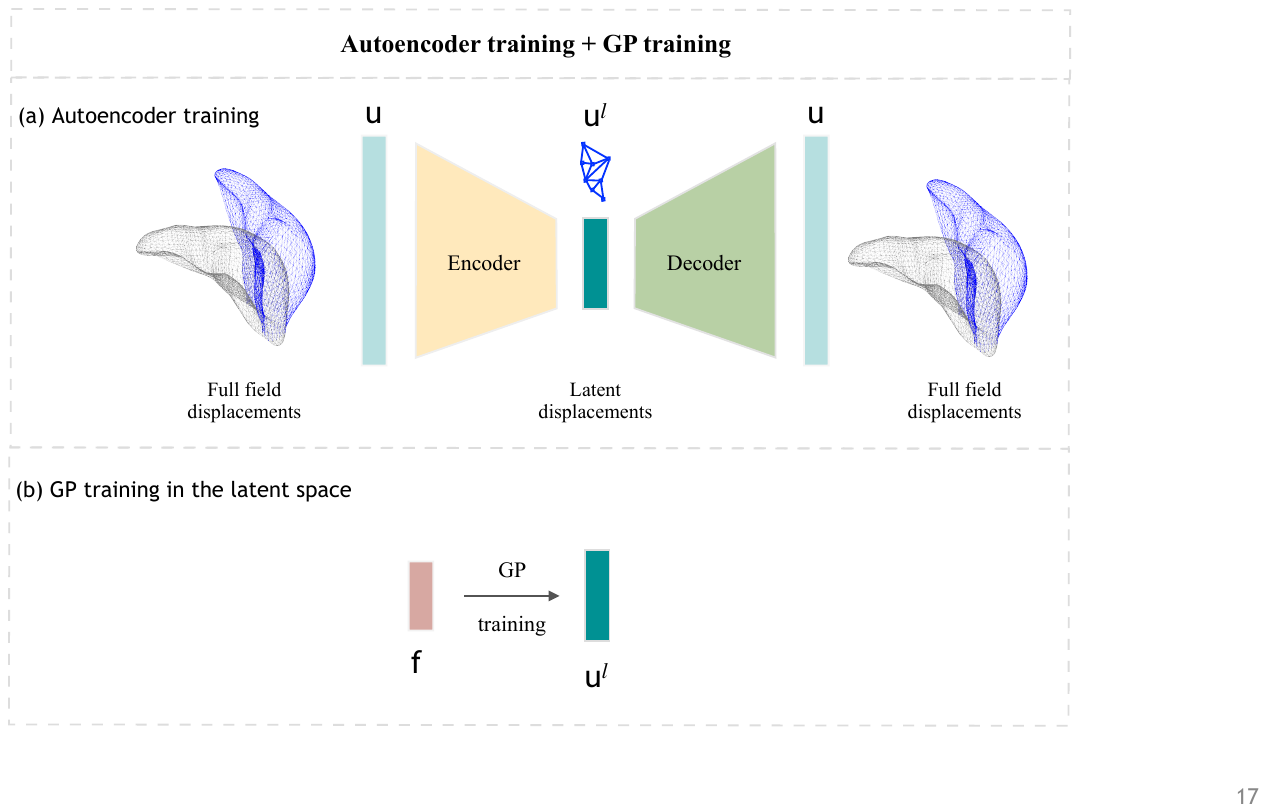}
     \caption{Two-stage training. (a) First, the autoencoder neural network is trained to compress full field displacement data to corresponding latent space representations. (b) Afterwards, the GP model is trained to provide the probabilistic force-displacement mapping in the latent space.}
     \label{fig: offline}
\end{figure}

The training of the two constituent models is performed sequentially in two stages, see Figure~\ref{fig: offline}. In the first stage, Figure~\ref{fig: offline}a, the autoencoder network is trained only on full-field displacement data to get the corresponding compressed representations, $\bfm{u}^l$. In the second stage, Figure~\ref{fig: offline}b, the GP regression model is trained between the forces, $\bfm{f}$, and the latent displacements vector $\bfm{u}^l$.

\begin{figure}[t]
     \centering
     \includegraphics[width=0.8\textwidth]{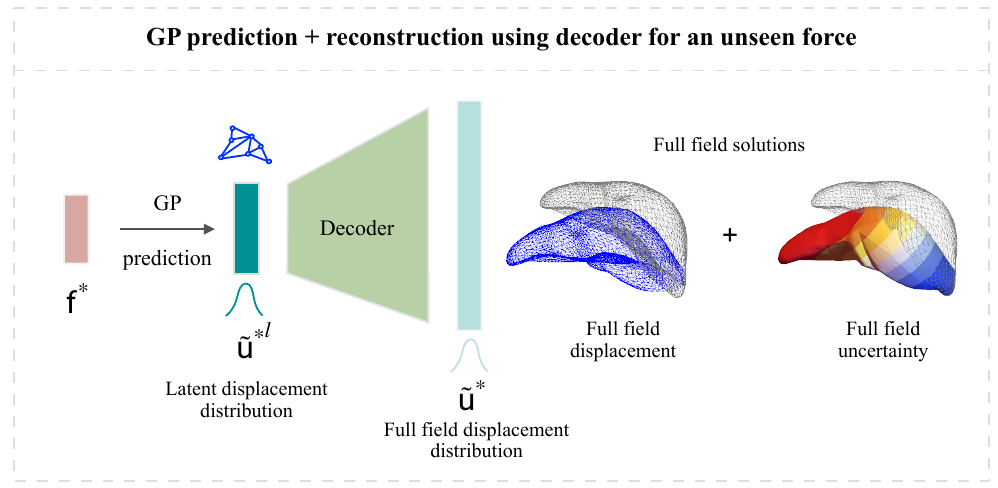}
     \caption{For an unseen input force $\mathbf{f}^*$, the GP is used to predict the latent displacement distribution (i.e., our uncertainty about the displacements is described by a probability distribution). Subsequently, these latent displacements are projected to the full-field displacement distribution by using the decoder component of the autoencoder network.}
     \label{fig: online}
\end{figure}

Following the training, the GPR model is stacked with the Decoder part of the autoencoder network, which provides the probabilistic force-full-displacement model, see also Section~\ref{sec: latent to full}. Specifically, for a given unseen input force, GPR model predicts the latent displacement distribution. These latent displacements are then reconstructed to the full space using the decoder part of the autoencoder network, see Figure~\ref{fig: online}.

\subsection{Obtaining latent representations with autoencoder neural networks}
\label{sec: autoencoder model}

Autoencoders are neural networks that are primarily used for unsupervised learning and dimensionality reduction tasks~\citep{auto_hinton}. They are designed to encode high-dimensional input data into a lower-dimensional latent space representation, and then decode it to the original input space, aiming to reconstruct the input data as accurately as possible. 

The architecture of an autoencoder network typically consists of two main components: an encoder and a decoder. The encoder takes the input data and maps it to a compressed representation in the latent space, while the decoder takes this compressed representation and reconstructs the original input data. The encoder and decoder are usually symmetrical, meaning that they have the same number of layers and the layer sizes are mirrored. One of the key aspects of autoencoders is that the latent space, also referred to as the bottleneck layer, has a lower dimensionality than the input space. This forces the network to learn a more efficient and meaningful representation of the data. By compressing the data into a lower-dimensional space and then reconstructing it, autoencoders can capture the most salient features by discarding less relevant information.

The autoencoder network, $\mathcal{U}$, is constructed by subsequently applying the encoder and decoder networks 
\begin{equation}
      \mathcal{U}(\bfm{u},\bm{\theta}_{\text{auto}})=\mathcal{U}_{\text{decoder}}(\mathcal{U}_{\text{encoder}}(\bfm{u},\bm{\theta_{\text{auto}}}),\bm{\theta_{\text{auto}}}), 
\end{equation}
where $\bfm{u}$ stands for the full field displacements and  $\bfm{u}^{l}=\mathcal{U}_{\text{encoder}}(\bfm{u},\bm{\theta_{\text{auto}}})$ is their corresponding latent/compressed representation. For a given full field displacement dataset, $\{\bfm{u}_i\}_{i=1,..,N}$, the autoencoder network
is trained by minimizing the following mean squared error loss 
\begin{equation}
      \mathcal{L}(\mathcal{D}^{f},\bm{\theta}_{\text{auto}}) = \frac{1}{N}\sum_{i=1}^{N} \norm{\mathcal{U}(\bfm{u}_i,\bm{\theta}_{\text{auto}})-\bfm{u}_i}_2^{2}, \label{eq:lossDeterm}
\end{equation}
where $\bm{\theta}_{\text{auto}}$ are trainable parameters of the autoencoder network. The optimal parameters, $\bm{\theta}_{\text{auto}}^{*}$, are retrieved by minimizing the loss function:
\begin{equation}
      \bm{\theta_{\text{auto}}}^{*} = \text{arg }\underset{\bm{\theta_{\text{auto}}}}{\text{min}}~\mathcal{L}(\mathcal{D}^{f},\bm{\theta_{\text{auto}}}). \label{eq:minimisationLossDeterm}
\end{equation}
Therefore, once the autoencoder network is trained, for a given full field displacement field  $\bfm{u}$, the corresponding latent representation $\bfm{u}^{l}$ is computed by applying the encoder part of the autoencoder network: 
\begin{equation}
    \bfm{u}^{l} = \mathcal{U}_{\text{encoder}}(\bfm{u},\bm{\theta_{\text{auto}}}^{*}).
    \label{eq: encoded}
\end{equation}
Hence, we obtain latent state dataset $\mathcal{D}^{l}=\{(\bfm{f}_i,\bfm{u}^{l}_i)\}_{i=1,..,N}$, from the full field dataset  $\mathcal{D}^{f}=\{(\bfm{f}_i,\bfm{u}_i)\}_{i=1,..,N}$. Note that forces are identical for both datasets and they are always provided as smaller dimensional arrays, which is discussed in detail in Section~\ref{sec: data_generation}. In the next section, we will use the latent dataset, $\mathcal{D}^{l}$, to elaborate Gaussian process regression model, which will provide the transformation between the input forces, $\bfm{f}$, and the latent solutions, $\bfm{u}^l$. 

\emph{Remark:} In our study, we employ two distinct autoencoder architectures, see Section~\ref{sec: examples DNN architectures}. The first type utilizes fully connected networks, suitable for handling arbitrary unstructured meshes. Meanwhile, the second type utilizes convolutional neural networks (CNNs) specifically designed for structured mesh scenarios \citep{cotin, DESHPANDE2022115307}.

\subsection{Gaussian Process Regression in the latent space}\label{sec: GP}

As discussed in the introduction, Gaussian Process Regression (GPR)
faces scalability challenges as the number of data points, $N$, and the dimension of outputs, $M$, increase. In this work we focus on reducing the bottleneck related to $M$, which is particularly important for problems with a large number of degrees of freedom, such as the examples studied in Section~\ref{sec: results}. As explained earlier, the main idea behind reducing $M$ is to project the high-dimensional displacements (of dimension $M$) to a significantly lower-dimensional space, $L \ll M$, and then employ GPR between the input forces and the latent representation of displacements.
In addition to that, in this work we follow a common practice to implement independent GPs for each dimension of the output, assuming that all output components are independent of each other. This approach significantly reduces the $\mathcal{O}(N^3L^3)$ complexity of a multi-output GPR problem to $L$ independent problems of $\mathcal{O}(N^3)$ complexity each (i.e., $\mathcal{O}(N^3L)$). Because our framework uses independent GPRs for each latent output, in this subsection, we provide a brief introduction to single-output GPs. For more details, interested readers are referred to \citep{C_Rasmussen_2006}.

GPs are an extension of multivariate normal distributions into an infinite-dimensional Gaussian distribution \citep{A_Gelman_2003}. Consider two data points $(\bfm{f},w(\bfm{f})), (\bfm{f}',w(\bfm{f}'))$, a GP is completely specified by its mean function and covariance function. We define mean function $m(\bfm{f})$ and the covariance function $k(\bfm{f},\bfm{f}')$ of a real process $w(\bfm{f})$ as: 
\begin{equation}
    \begin{split}
        m(\bfm{f}) &= \mathbb{E}(w(\bfm{f})), \\
        k(\bfm{f},\bfm{f}') &= \mathbb{E}\left((w(\bfm{f}) -m(\bfm{f}) )(w(\bfm{f}') -m(\bfm{f}') )\right).
    \end{split}
\end{equation}

As mentioned, a GP is used to describe a distribution over functions. Therefore, a realization of a GP reads as:
\begin{equation}
\label{Eq: GP realization}
w(\bfm{f}) \sim \GP(m(\bfm{f}),k(\bfm{f},\bfm{f}')),
\end{equation}
where $\bfm{f}$ and $\bfm{f}'$ are two GP input vectors of dimension $D$, which are force vectors in the scope of the present framework. Hence, in general, for a finite collection of inputs $\bfm{F}=[\bfm{f}_1, \cdots, \bfm{f}_N]$: 
\begin{equation}
\label{finite dimensional}
\bfm{w} \sim \N(\bfm{m},\bfm{K}),
\end{equation}
where $\bfm{w}=\begin{bmatrix} w_{1}&\cdots&w_{N}\end{bmatrix}^{T}$, $\bfm{m}=\begin{bmatrix} m({\bfm{f}}_{1})&\cdots&m({\bfm{f}}_{N})\end{bmatrix}^{T}$ and $\bfm{K}$ denotes the covariance matrix between two inputs, i.e., $(\bfm{K})_{ij}=k(\bfm{f}_{i},\bfm{f}_{j})$. Frequent choices for the covariance function are given in \citep{C_Rasmussen_2006}.

In simple terms, in GPR, we set a GP (Eq.\eqref{Eq: GP realization}) as a prior for the function that maps inputs to outputs and then update this prior with available observations using Bayes' rule (see \cite{hus}). In our framework, the GP predicts displacements corresponding to latent space representations, as shown in Figures\ref{fig: offline} and \ref{fig: online}. Therefore, we will use the superscript-$l$ notation, $(\cdot)^l$, for all GP outputs.

Since the proposed framework uses independent GPRs for latent outputs, we will focus on a single latent output to explain the GP setup. Let $\tilde{u}^{l}_{i}$ be an observation for a particular latent output contaminated by Gaussian noise, i.e., $\tilde{u}^{l}_{i} \sim \N(w(\bfm{f}_{i}), \sigma^{2})$, with $i = 1, \cdots, N$. Without loss of generality, we assume that the function mapping inputs to the output is a realization of a GP with zero mean and a covariance function $k(\bfm{f}, \bfm{f}')$, defined by its parameter set $\bm{\theta}_{\text{GP}}$. Furthermore, let $\bfs{y}=[\tilde{u}^{l}_{1}, \cdots, \tilde{u}^{l}_{N}]$. The predictions ${\tilde{u}}^{l*}$ of a GPR for a new data point input $\bfm{f}^{*}$ will be as follows:
\begin{equation}
\label{GPR Distribution}
{\tilde{u}}^{l*}|\bm{\theta}_{\text{GP}}, \sigma^{2},\tilde{\bfm{u}}^{l},\bfm{F},\bfm{f}^{*} \sim \N(\mathbb{E}({\tilde{u}}^{l*}),\mathbb{V}({\tilde{u}}^{l*})),
\end{equation}
\begin{equation}
\label{GPR mean}
\mathbb{E}({\tilde{u}}^{l*})=\bfm{k}^{*T}(\bfm{K}+\sigma^2\bfm{I}_{N})^{-1}\bfs{y},
\end{equation}
and 
\begin{equation}
\label{GPR variance}
\mathbb{V}({\tilde{u}}^{l*})=k(\bfm{f}^{*},\bfm{f}^{*})-\bfm{k}^{*T}(\bfm{K}+\sigma^2\bfm{I}_{N})^{-1}\bfm{k}^{*},
\end{equation}
where $\mathbb{E}({\tilde{u}}^{l*})$ denotes mean of the predicted $\tilde{u}^{l*}$ at $\bfm{f}^{*}$, $\mathbb{V}({\tilde{u}}^{l*})$ denotes the prediction variance, $\bfm{I}_{N}$ is an $N \times N$ identity matrix, $\bfm{K}$ denotes the $N \times N$ covariance matrix given in Eq.~\eqref{finite dimensional}, $\bfm{k}^{*}$ is an $N \times 1$ vector with $(\bfm{k}^{*})_i = k(\bfm{f}^{*},\bfm{f}_{i})$. Furthermore, to obtain the optimal parameters for GPR, we train the GP by maximizing the following equation:
\begin{equation}
\label{marginal likelihood}
\log p(\bfs{y}|\bfm{F},\bm{\theta_{\text{GP}}},\sigma^{2})=-\frac{1}{2}\bfs{y}^{T}(\bfm{K}+\sigma^{2}\bfm{I}_{n})^{-1}\bfs{y}-\frac{1}{2}\log[\text{det}(\bfm{K}+\sigma^{2}\bfm{I}_{n})]-\frac{n}{2}\log(2\pi).
\end{equation}

In Equation \eqref{marginal likelihood}, $p(\bfs{y}|\bfm{F},\bm{\theta_{\text{GP}}},\sigma^{2})$ represents the conditional likelihood of the observations. This likelihood depends on the parameter set $\bm{\theta_{\text{GP}}}$, the input matrix $\bfm{F}$, and the noise variance $\sigma^{2}$. The optimised parameters ($\bm{\theta_{\text{GP}}}$, $\sigma^{2}$) are computed by maximising the log likelihood, as discussed in \cite{hus}. Note also that the GP and the autoencoder network are trained separately. Autoencoder network needs to be trained first in order to generate the latent state dataset, which is then used to train the GP.

\subsection{Projecting latent GP predictions to the full field space using decoder} \label{sec: latent to full}

As depicted in the Figure~\ref{fig: online}, the latent space predictions obtained using GP are projected to the full space using decoder part of the autoencoder network. At the inference stage, for a new input force $\bfm{f^{*}}$, the GPR first generates latent displacement distribution $p({\tilde{u}}^{*l}|\bfm{f^{*}})$ for each latent output, $l$, which is a Gaussian distribution with the mean and variance obtained using Eq.(\ref{GPR mean}) and Eq.(\ref{GPR variance}) respectively:

\begin{equation}
    p({\tilde{u}}^{*l}|\bfm{f^{*}}) = \mathcal{N}(\bfm{\mathbb{E}}_{\text{GP}}, \bfm{\mathbb{V}}_{\text{GP}}).
\end{equation}

Subsequently, the distributions of full field displacements are obtained by translating samples from the latent predictive distributions through the decoder component. Let the latent vector $\bfm{u}_{s}^{l}$ denote a sample, $s$, drawn from distributions corresponding to all latent outputs, obtained using independent GPs for an input $\bfm{f}^{*}$. The displacement corresponding to the full field solution for each degree of freedom (DOF) is a Gaussian distribution, whose mean and variance are computed as follows:
\begin{equation}
  \begin{split}
    (\tilde{u}^{*}_{\mu})_i & \approx \frac{1}{S} \sum_{s=1}^{S} (\mathcal{U}_{\text{decoder}}(\bfm{u}_{s}^{l}))_i,  \\
    (u_{\sigma}^{*})_i^{2} & \approx \frac{1}{S-1} \sum_{s=1}^{S} ((\mathcal{U}_{\text{decoder}}(\bfm{u}_{s}^{l}))_i -  (u^{*}_{\mu})_i)^2,
   \end{split}
   \label{eq: mean_std_prediction}
\end{equation}
where $i$ represents index of DOF. In order to get the full field displacement distributions, we generate $S$ samples in the latent space, and these samples are then projected to the original space using the decoder. Now the final Gaussian distributions of displacements of the full field solution is represented through the mean ($\tilde{\bfm{u}}^{*}_{\mu}$) and standard deviation (${\bfm{u}_{\sigma}^{*}}$) of these $S$ samples. We set $S=300$ for all the implementations presented in this work.

\subsection{Finite element formulation for non-linear deformations of solid bodies} \label{sec: fem_description}

We consider the problem of deformation of a body occupying the domain $\Omega$, that is fixed at a part of its boundary $\Gamma_{\text{u}}$ (Dirichlet boundary conditions), and loaded over the surface $\Gamma_{\text{t}}$ (Neumann boundary conditions) and/or by body forces $\bfs{\bar{b}}$. This boundary value problem is expressed by the virtual work principle
  \begin{equation}
    \label{eq:VWP}
          \int_{\Omega} \bfs{P}(\bfs{F}(\bfs{u})) \cdot \nabla\delta\bfs{u}\, \rd V - \int_{\Omega} \rho\,\bfs{\bar{b}} \cdot \delta\bfs{u}\, \rd V -
      \int_{\Gamma_t} \bfs{\bar{t}} \cdot \delta\bfs{u}\, \rd S   = 0 \quad\quad\forall\delta\bfs{u},
  \end{equation}
where $\bfs{u}$ and $\delta\bfs{u}$ belong to appropriate functional spaces, $\bfs{u}=\bar{\bfs{u}}$ and $\delta\bfs{u}=\bfm{0}$ on $\Gamma_{\text{u}}$, and $\bfs{P}(\bfs{F})$ is the first Piola-Kirchhoff stress tensor. $\bfs{P}(\bfs{F})$ is given by the constitutive relationship through the strain energy potential, $W(\bfs{F})$, as:
  \begin{equation}
    \label{eq:const}
   \bfs{P}(\bfs{F}) = \frac{\partial W(\bfs{F})}{\partial \bfs{F}},
  \end{equation}
where $\bfs{F}=\bfs{I} + \nabla \bfs{u}$ is the deformation gradient tensor. In this work we use following Neo-hookean (hyperelastic) strain energy density: 
\begin{equation}
 W(\bfs{F})=\frac{\mu}{2}(I_c-3-2 \ln{J})+\frac{\lambda}{4}( J^2-1-2 \ln{J}),
 \label{Eq:NeoHoohe}
\end{equation}
where $J = \text{det}(\bfs{F})$ and $I_{\text{c}} = \text{tr}(\bfs{F}^{T}\bfs{F})$. 

Following conventional finite element methodology, the problem represented by Eq.~(\ref{eq:VWP}) is discretized and posed as a system of nonlinear equations for the unknown nodal displacements, which is then solved iteratively using Newton-Raphson method. As the effect, for an input vector of prescribed external forces, the vector of nodal displacements is computed. The solutions for different input forces provide the necessary force-displacement dataset, in which each element consist of the pair $(\bfm{f},\bfm{u})$, where $\bfm{f}$ encodes the information on external loading and $\bfm{u}$ is the vector of full-field displacement of FE mesh. 

\emph{Remark:} The vector $\bfm{f}$ provides a compressed information about the external loading. In our benchmark examples, a body can be subjected to either point loads applied to a single node, or uniform body forces acting throughout the entire volumes. In the 2D example, we apply point loads, and describe the loading $\bfm{f}$ as its magnitude ($f_x$, $f_y$) and the position, $d$, at which the force is applied (relative to the fixed boundary). In the 3D example, we apply body forces, and the loading $\bfm{f}$ represents the force density components ($b_x$, $b_y$, $b_z$). See also more details in Section~\ref{sec: data_generation}.


\section{Results}
\label{sec: results}

\begin{figure}[!ht]
     \centering
     \subfloat[]{\includegraphics[width=0.35\textwidth]{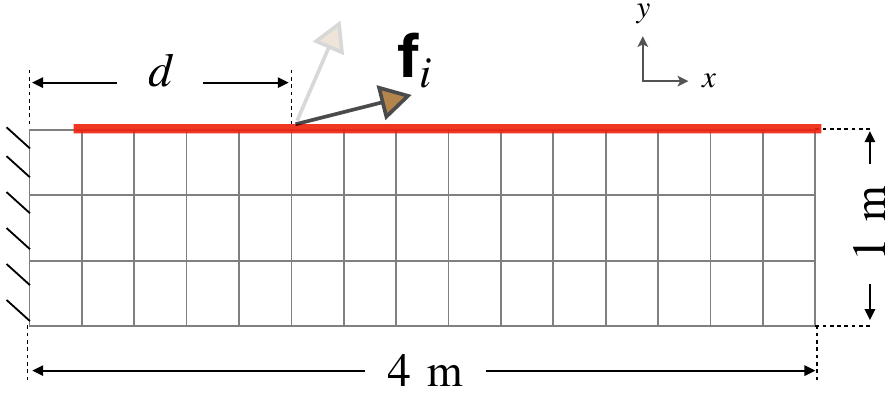}\label{fig: beam_schematic}}\hspace{0.1\textwidth}
     \subfloat[]{\includegraphics[width=0.3\textwidth]{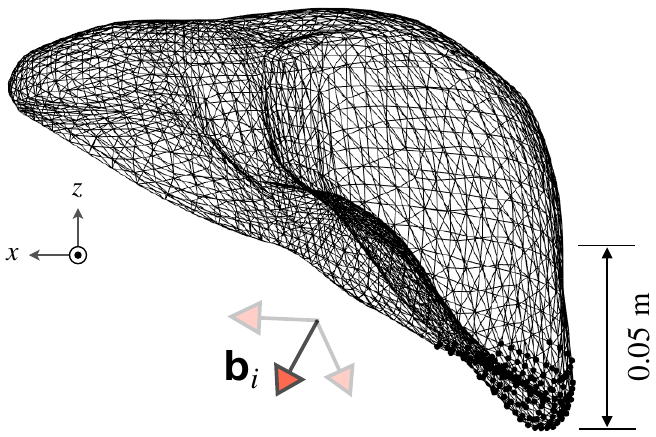}\label{fig: liver_schematic}}
      \caption{Schematics of examples considered in this work. (a) 2D beam discretised with quad elements is subjected to point loads on the nodes lying on red line; this example is adapted from \citep{DESHPANDE2022115307}. (b) 3D liver is subjected to body forces.  }
     \label{fig: schematic_data}
\end{figure}

The performance of the framework is demonstrated and studied based on two benchmark examples, as illustrated in Figure~\ref{fig: schematic_data}: the 2D beam
example and the 3D liver example. The bodies are subjected to varied loading conditions and undergo large deformations in non-linear regime. The data necessary for training is generated through the finite element simulations, see  Section~\ref{sec: fem_description}, and the procedure and data range is described in a more detail in Section\ref{sec: data_generation}. \rev{The finite element simulations are performed with the AceFEM framework~\citep{acegen, korelc2016automation}.} The details of the surrogate framework and its training are provided in Section~\ref{sec: implementation details}. In Sections~\ref{sec: sources of errors}-\ref{sec: missing data region} we study the predictive capabilities of the framework, with a more detailed discussion on sources of errors and on the interpretation of probabilistic predictions. \rev{In particular, in Sec. \ref{sec: missing data region}, we demonstrate the capabilities of our framework to be a convenient indicator of the missing data region.}

\subsection{Dataset generation information}\label{sec: data_generation}

In the 2D beam example, see Figure~\ref{fig: beam_schematic}, a beam is loaded with point forces of random direction and magnitude, within the ranges described in Table~\ref{tab:datasets}. These forces are applied to the nodes located along the top line, which is visually represented by red line in the Figure~\ref{fig: beam_schematic}. For the 2D beam case, displacement data is provided for all degrees of freedom (DOFs) of the mesh, capturing the full-field displacements of the beam. However, the forces used as inputs to the surrogate framework are represented in a compressed format of a 3-component array. Specifically, the applied point load is characterized by two components, namely $f_x$ and $f_y$, representing the force magnitudes in the x and y directions, respectively. Additionally, the distance, $d$, from the fixed boundary to the point of application is considered. This information is visually depicted in Figure~\ref{fig: beam_schematic}.

In the 3D liver example, a body representing the liver geometry is subjected to a body force of random direction and magnitude. The liver is also constrained at specific nodes located at the right end, see Figure~\ref{fig: liver_schematic}. Similarly to the beam case, displacement data is provided for all DOFs of the liver mesh, representing the full-field displacements of the organ. The compressed representation of forces consists of the body force density components in the x, y, and z directions, respectively ($b_x, b_y, b_z$). Further details, such as the range of external forces applied, as well as other information used to generate the datasets are given in the Table~\ref{tab:datasets}.

\rev{\noindent\emph{Remark:} Note that for each loading case, the data generated by the FE simulation is deterministic. This has been a choice in the scope of the present work, however, it can be extended to the non-deterministic case if needed.}

\begin{table}[h!]
    \small
    \centering
    \begin{tabular}{l|R|R|L|R}
        Problem & N.of DOFs ($\mathcal{F})$ & Range (External forces/ body force density)& Young's modulus $E$ [\text{Pa}], Poisson's ratio  $\nu$, density  $\rho$ [kg/m$^3$] & Dataset size (train+test)\\
        \hline
        2D beam & 128 & $f_x, f_y$ = -2.5 to 2.5 N & 500, 0.4, - &$5700 + 300$ \\
        3D liver & 9171 & $b_x, b_y, b_z$= -0.8 to 0.8 $\text{N}/\text{kg}$ & 5000, 0.45, 1000
    &$7600 + 400$ 
    \end{tabular}
    \caption{Details of FE datasets.}
    \label{tab:datasets}
\end{table}

\subsection{Implementation details}
\label{sec: implementation details}

\subsubsection{Autoencoder networks:  Architectures \& training details}
\label{sec: examples DNN architectures}

\begin{figure}[h!]
     \centering
     \includegraphics[width=0.95\textwidth]{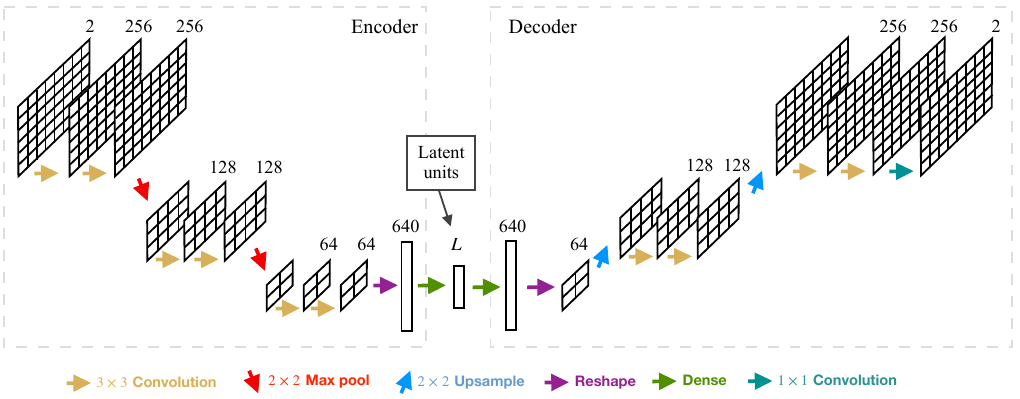}
     \caption{CNN autoencoder architecture for used for the 2D beam case. Input is provided to the network in the structured mesh format, the number at the top of the convolution layer output indicates the number of channels in the respective layer. Input is applied series of convolution and pooling/upsampling layers until the original dimension is retrieved. In the bottleneck level, fully connected layers (dense layer) are applied and the number the top of their output represents the number of units present in it. }
     \label{fig: 2dautoencoder_architecture}
\end{figure}

\begin{figure}[h!]
     \centering
     \includegraphics[width=0.95\textwidth]{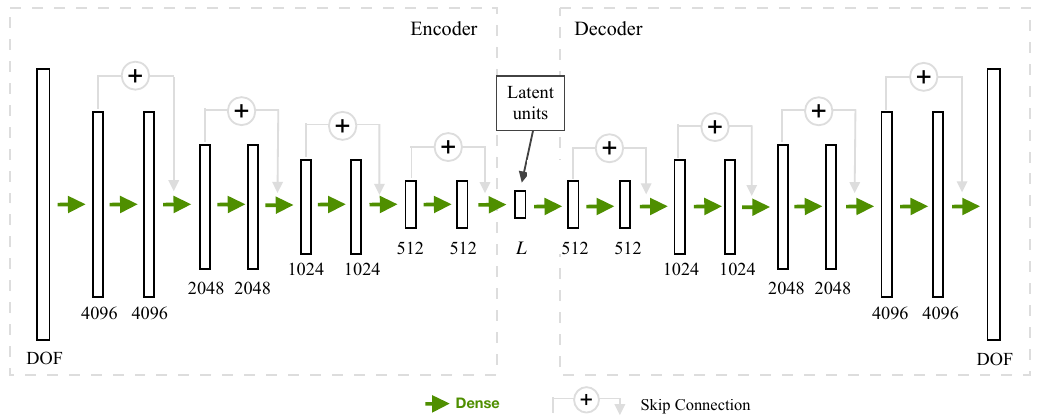}
     \caption{Fully connected autoencoder architecture used for the liver case. Skip connections are indicated with the (+) sign. The number of units for each dense layer output is indicated at the bottom. The latent representation dimension is denoted as $L$, set to $L = 16$ for the 3D liver case.}
     \label{fig: fc_architecture}
\end{figure}

In this study, we employ two types of autoencoder networks. Firstly, we utilize a CNN-based architecture, illustrated in Figure~\ref{fig: 2dautoencoder_architecture}, tailored for the 2D beam example. Secondly, we utilize a fully connected autoencoder network, depicted in Figure~\ref{fig: fc_architecture}, implemented for the 3D liver example. Both of these networks consist of an encoder and a decoder component. The primary function of these autoencoder networks is to compress the displacement solutions of the full field space into corresponding latent representations. Note that the forces are initially provided in a simple format, see Sec.~\ref{sec: data_generation}, thus do not need to be compressed by autoencoder networks.

\textbf{Architecture for the 2D beam case}\\
Encoder: The input to the CNN autoencoder is a mesh displacement tensor, which is represented with 2 channels. These channels correspond to the displacements along the x and y DOFs of the mesh. The input tensor undergoes application of two convolutional layers, each with 256 channels and $3\times3$ filters as shown in Figure~\ref{fig: 2dautoencoder_architecture}. Convolutional operations enable feature extraction and dimensionality reduction. Subsequently, a max-pooling operation is applied, which reduces spatial dimensions of the tensors while preserving the number of channels. This process is repeated again with 128 channels, further capturing important patterns in the data. At the latent level of the network, two convolution layers with 64 channels are applied, which is followed by flattening of the tensor. Now the flattened tensor is applied with a dense layer with $l$ units, which stands for the dimension of the compressed/latent state. This compressed representation contains crucial information about the input mesh, capturing the most relevant features.

Decoder: The Decoder takes a flattened tensor of dimension $L$ as its input. Initially, this input tensor undergoes processing via a dense layer. This strategic utilization of the dense layer facilitates a transformation of the arbitrary latent dimensions into a new size that can be conveniently reshaped into a grid structure, a format that seamlessly aligns with the subsequent convolutional layer operations. For instance, utilizing a dense layer of 640 units as shown in the Figure~\ref{fig: 2dautoencoder_architecture} allows us to reshape the flat tensor to a $64\times5\times2$ tensor, on which series of upsampling and convolution operations are performed until the original spatial dimensions is achieved. In the end, a 2 channel convolution with $1\times1$ filter is applied to get back the original mesh shape. 

\textbf{Architecture for the 3D liver case} \\ 
Convolutional neural networks are inherently designed to work with grid-like inputs, such as structured meshes. For that reason, this methodology isn't immediately adaptable to inputs that lack this grid-based nature, such as the unstructured mesh that are present in the liver example. To address this limitation, we introduce an alternative solution in the form of a fully connected autoencoder network as illustrated in Figure~\ref{fig: fc_architecture}. This network can handle diverse and unstructured meshes. 

Encoder: This component receives input in the form of displacements linked to all degrees of freedom (DOFs) of the mesh. This information is encoded within a flattened tensor, which serves as the input layer. It is further applied with two dense layers with 4096 units each. We also introduce skip connections as illustrated in the Figure~\ref{fig: fc_architecture}, which avoid vanishing gradient issues and stabilize the training procedure. This procedure applying blocks of dense layers and skip connections is repeated three more times with 2048, 1024, 512 units for dense layers respectively. 

Decoder: Decoder takes the $L=16$ dimensional 1D tensor as the input (compressed/latent tensor), a similar procedure of applying two dense layers and skip connections is followed until the original size tensor is obtained. The number of units used in each dense layer is detailed in the Figure~\ref{fig: fc_architecture}.

\textbf{Training of autoencoder networks}\\
Autoencoder network is trained on the full field displacement dataset, using the loss function as described by Eq.(\ref{eq:lossDeterm}). Minimization is carried out using the Adam optimizer with recommended parameter configurations as presented by Kingma \cite{kingma2017adam}. In both instances, a batch size of 16 is employed, with the 2D beam case trained across 32000 epochs and the 3D liver case for 4000 epochs. The initial learning rate is set at 1e-4 and linearly decays to 1e-6 during the training process. Autoencoder networks are trained using Tensorflow~\citep{tfp} library on a Tesla V100-SXM2 GPU, utilizing the high-performance computing facilities at the University of Luxembourg \citep{ULHPC}.   

\rev{\noindent\emph{Remark:} Besides the size of the latent vector, there are several other hyperparameters that determine performance of both autoencoder architectures. In the case of 2D CNN network, these are: the size of convolution window, the number of channels per layer, the number of convolutions per level, and the type of pooling layers. In the case of 3D fully connected network, these are: the sizes of each layer, and the arrangement of skip connections. The particular choice of hyperparameters that determine both autoencoder architectures is primarily based on our earlier experience and studies, see \cite{DESHPANDE2022115307, DESHPANDE2024108055}, which is then followed by the grid search optimization in the space of hyperprameters. The presented final architectures have been found to perform the best for the problems analyzed in this work.}

\subsubsection{Gaussian process details}
All the Gaussian Process (GP) implementations presented in this work are performed using \rev{the Mat\'{e}rn covariance function, as it is known to be less prone to overfitting. (Note, however, that our framework is not limited to any particular covariance function.) The Mat\'{e}rn covariance function has the following form:
\begin{equation}
C_{\nu}(r_{ij}) = \lambda^{2}\frac{2^{1-\nu}}{\Gamma(\nu)} \left( \sqrt{2\nu} \frac{r_{ij}}{\psi} \right)^{\nu} K_{\nu}\!\left( \sqrt{2\nu} \frac{r_{ij}}{\psi} \right),
\end{equation}
where $r_{ij}=\lVert\mathbf{f}_{i}-\mathbf{f}_{j}\rVert$, $\mathbf{f}_{i}$ and $\mathbf{f}_{j}$ denote the $i^{\text{th}}$ and $j^{\text{th}}$ input vectors, $\lVert\cdot\rVert$ denotes the Euclidean norm, $\psi$ is the length scale, $\Gamma(\nu)$ denotes the gamma function, $K_{\nu}$ is a modified Bessel function, $\lambda^{2}$ denotes the variance. The positive parameter $\nu$ is critical; the larger it is, the smoother $w(\bfm{f})$ becomes in Eq.~(\ref{Eq: GP realization}). A Gaussian Process (GP) with the Matérn covariance function is $\lceil \nu \rceil-1$ times mean square differentiable, and for $\nu \to \infty$, the Matérn covariance function converges to the radial basis function (RBF). If we choose $\nu$ in the form of a half-integer, i.e., $\nu = p + 1/2$, it leads to a convenient form of the covariance function that is a product of a polynomial of order $p$ and an exponential function. Often, as in this contribution, $\nu$ is fixed due to computational complexities and the extreme changes in the functions that occur as $\nu$ changes. The most common choices for $\nu$ are $1/2$, $3/2$, and $5/2$. In this work, we decided to use $\nu=5/2$ by comparing the predictions of the aforementioned covariance functions for the latent space. Accordingly, the $5/2$~Mat\'{e}rn covariance function reads:
\begin{equation}
\label{eq:kernel}
k(r_{ij}) = C_{5/2}(r_{ij}) = \lambda^{2}\Big(1+\frac{\sqrt{5}r_{ij}}{\psi}+\frac{5r_{ij}^2}{3\psi^2}\Big)\text{exp}\bigg(-\frac{\sqrt{5}r_{ij}}{\psi}\bigg).
\end{equation}
The remaining hyperparameters, $\psi$ and $\lambda^{2}$, are then inferred or optimized by maximizing the marginal likelihood in Eq.~(\ref{marginal likelihood}).  
}

The kernel hyperparameters are obtained by training GPs using the scikit-learn library \citep{pedregosa2011scikit} on the latent datasets, as described in Section~\ref{sec: GP}. Optimization is performed using the L-BFGS-B optimizer \citep{fletcher2000practical}. As described in Section~\ref{sec: GP}, we implement independent GPs corresponding to each latent component of the autoencoder network.

\subsection{Sources of prediction errors}
\label{sec: sources of errors}

\begin{figure}
    \centering
    \includegraphics[width=0.9\linewidth]{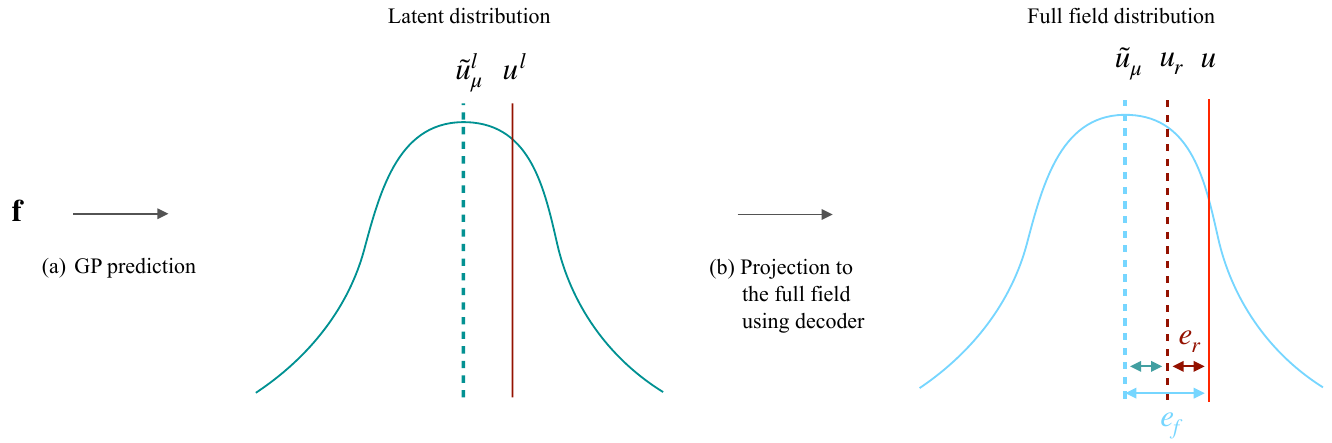}
    \caption{ (a) As described in the Figure~\ref{fig: online}, for an input force $\bfm{f}$, the latent distribution is predicted using GP. $\tilde{u}^{l}_{\mu}$ and $u^{l}$ represent the mean of the latent distribution and true latent solution for a particular latent DOF, respectively. (b) Latent distribution is then projected to the full space as described in Section~\ref{sec: latent to full}. $\tilde{u}_{\mu}$, $u_{r}$, $u$ represent the mean of the full field displacement prediction, reconstructed solution and the true FEM solution for a particular full field DOF, respectively. $e_\text{f}$ represents the framework error while $e_r$ represents the reconstruction error, for the particular full field DOF. }
    \label{fig: error_schematic}
\end{figure}

At the prediction phase, for a new unseen case, the GPR is first applied to achieve the latent space prediction, which is subsequently projected to the full space solution using the decoder part of the autoencoder network, see Section~\ref{sec: methodology} and Figure~\ref{fig: online}. Hence the error in the framework arises from GPR fitting as well as the compression and reconstructions steps done by the autoencoder network. The idea of this split of errors is schematically shown in Figure~\ref{fig: error_schematic} for a single DOF of the full-field solution.  

The mean prediction of the framework, $\tilde{u}_{\mu}$, is interpreted as the displacement solution. Hence the absolute difference between this mean prediction, $\tilde{u}_{\mu}$, and the true FEM solution, $u$, is considered as the error of the framework, 
\begin{equation}
   e_\text{f} = |\tilde{u}_{\mu} - u|.
   \label{eq: absolute error}
\end{equation}

The error contributed to autoencoder, $e_r$, is computed by taking absolute difference between the reconstructed solution and the FEM solution. The reconstructed solution is obtained by projecting the true latent solution (as obtained by Eq.~(\ref{eq: encoded})) to the full field, thus neglecting any contribution of the GP. The reconstructed solution reads $\bfm{u}_{r} = \mathcal{U}_{\text{decoder}}(\bfm{u}^{l})$, and the corresponding reconstruction error is 
\begin{equation}
   e_r = |u_{r} - u|.
   \label{eq: decode_error}
\end{equation}

Finally, the error contributed to GP fitting, $e_{\text{GP}}$, is computed by taking the difference between the mean full-field prediction, $\tilde{u}_{\mu}$, and the reconstruction solution, $u_{r}$, as follows: \begin{equation}
    e_{\text{GP}} = |\tilde{u}_{\mu} - u_r|.
    \label{eq: error_gp}
\end{equation}

Note: All the errors are computed in the decoded space, and are provided per full-field DOF.  

\subsection{Performance study of the GP+autoencoder framework}

\begin{figure}[t!]
     \centering
     
     \subfloat[Framework solution]{\includegraphics[width=0.47\textwidth]{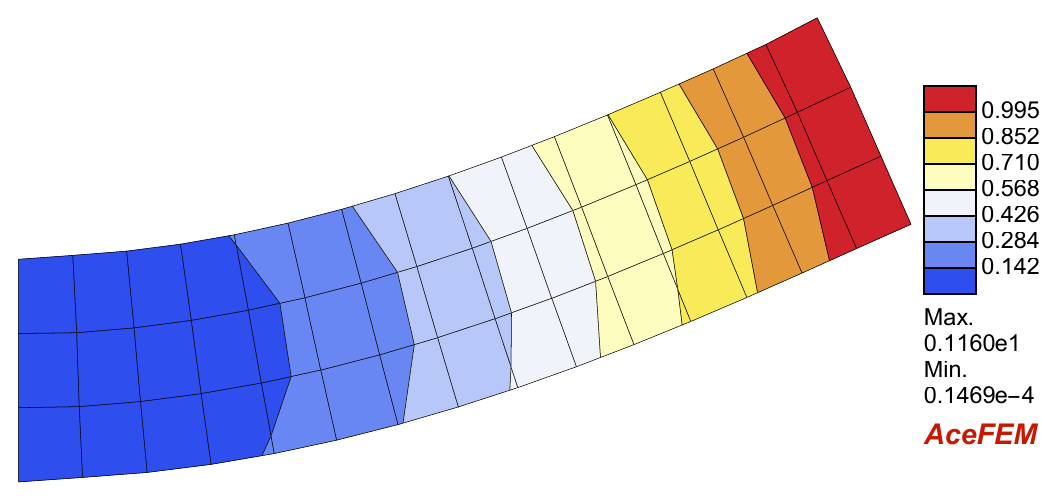}}\hspace{0.04\textwidth}
     \subfloat[FEM solution]{\includegraphics[width=0.465\textwidth]{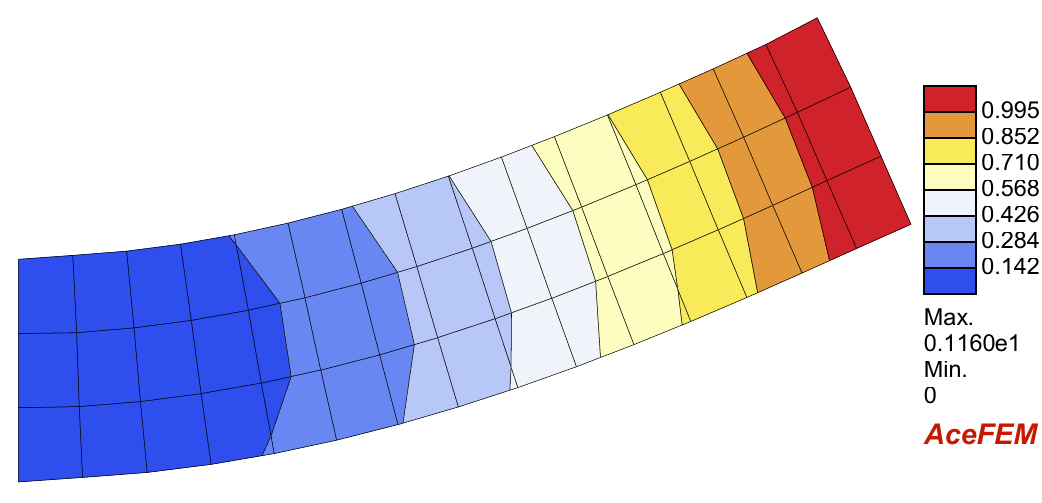}}
     
     \subfloat[Absolute error, $e_\text{f}$, see Eq.~\eqref{eq: absolute error}]{\includegraphics[width=0.47\textwidth]{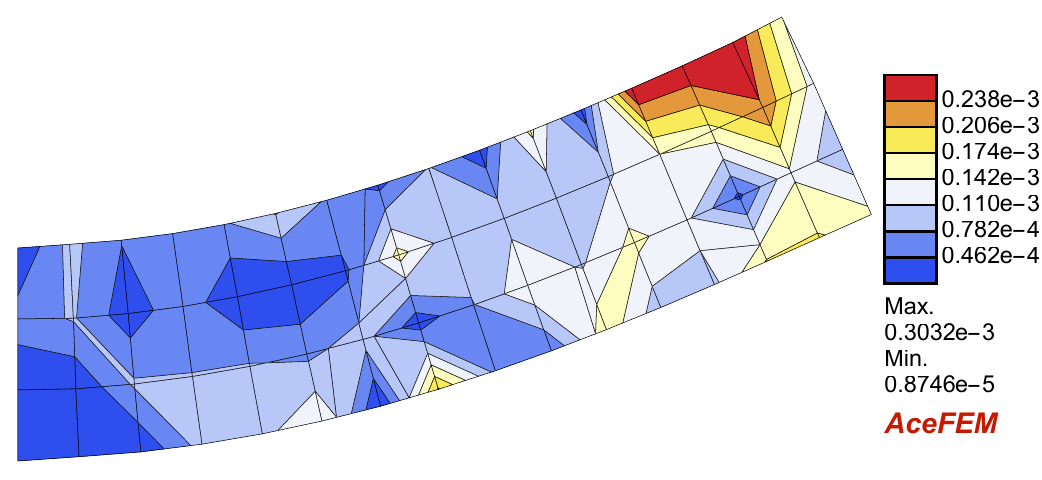}}\hspace{0.04\textwidth}
    \subfloat[Reconstruction error, $e_r$, see Eq.~\eqref{eq: decode_error}]{\includegraphics[width=0.47\textwidth]{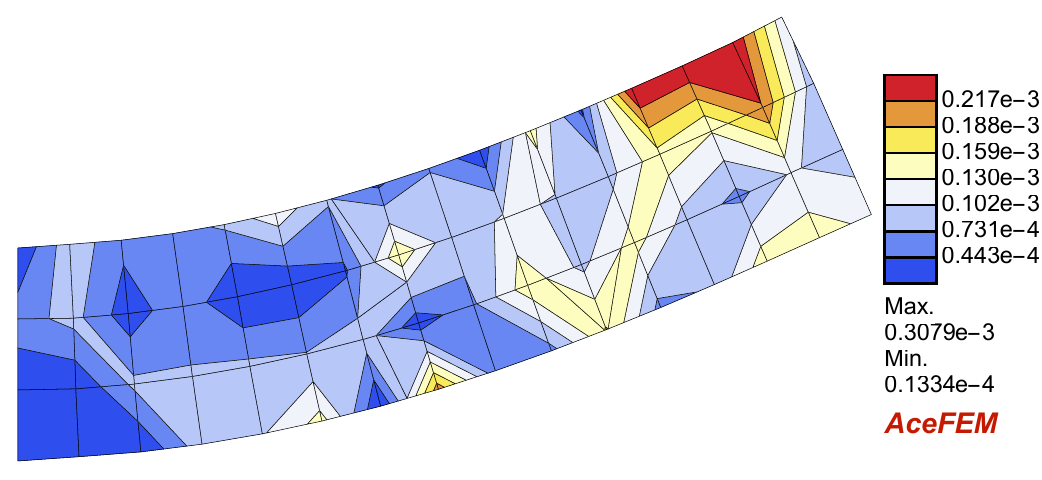}}\hspace{0.04\textwidth}
     
     \subfloat[GP component error, $e_{\text{GP}}$, see Eq.~\eqref{eq: error_gp}]{\includegraphics[width=0.47\textwidth]{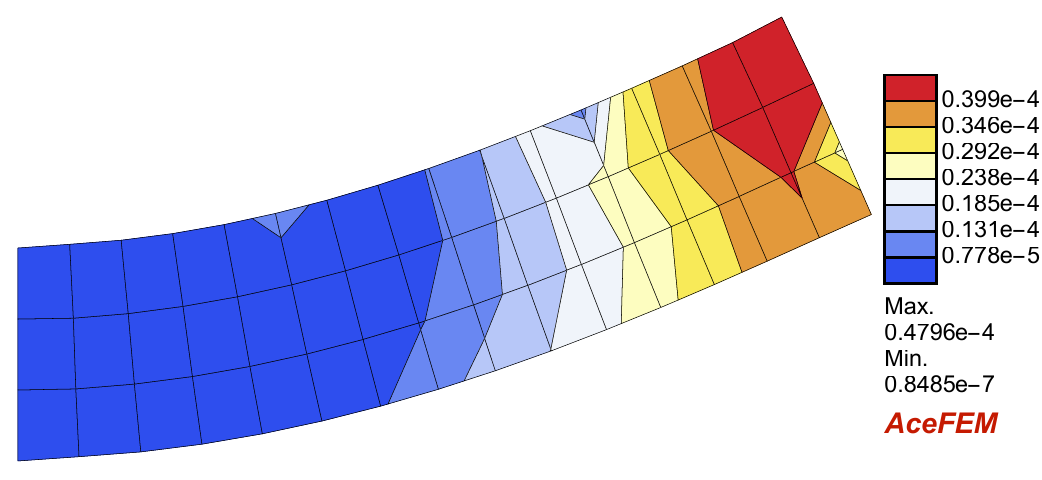}}\hspace{0.04\textwidth}
     \subfloat[Uncertainty prediction, $(u_{\sigma}^{*})^{2}$, see Eq.~\eqref{eq: mean_std_prediction}]{\includegraphics[width=0.47\textwidth]{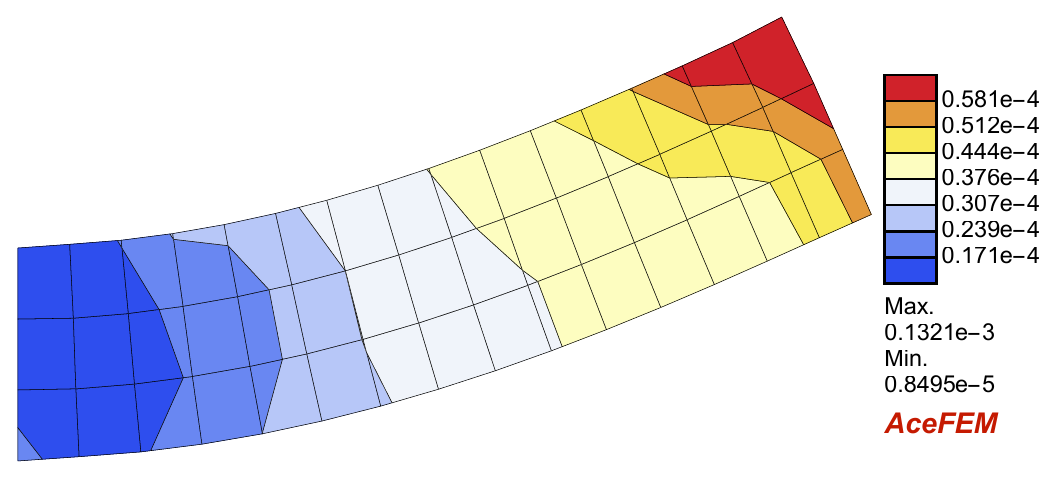}}
    
      \caption{Deformation of the 2D beam subjected to the point load (-2.00, 2.19) N on the top right corner node.  Nodal displacements obtained using the (a) proposed framework (b) FEM respectively. (c) Nodal error and (d) reconstruction error of the framework. (e) GP component error. (f) Uncertainty predicted by the framework for the full field solution. }
     \label{fig: 2D beam prediction}
\end{figure}

\begin{figure}[h!]
     \centering
     
     \subfloat[Framework solution]{\includegraphics[width=0.47\textwidth]{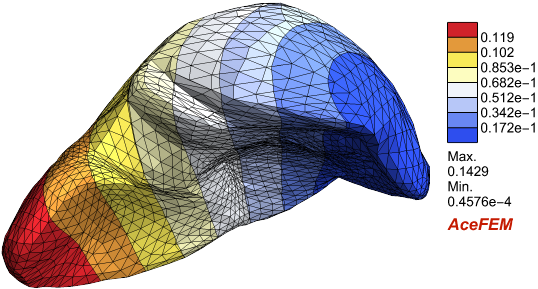}}\hspace{0.04\textwidth}
     \subfloat[FEM Solution]{\includegraphics[width=0.465\textwidth]{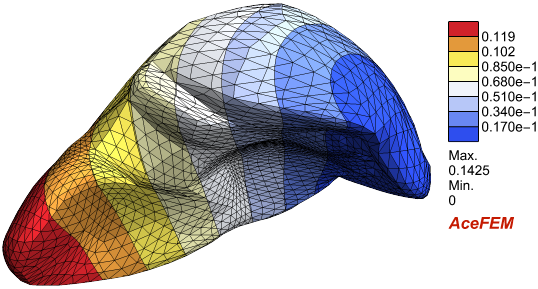}}
     
     \subfloat[Absolute error, $e_\text{f}$, see Eq.~\eqref{eq: absolute error}]{\includegraphics[width=0.47\textwidth]{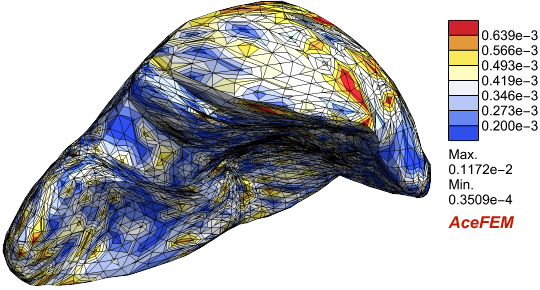}}\hspace{0.04\textwidth}
     \subfloat[Reconstruction error, $e_r$, see Eq.~\eqref{eq: decode_error}]{\includegraphics[width=0.47\textwidth]{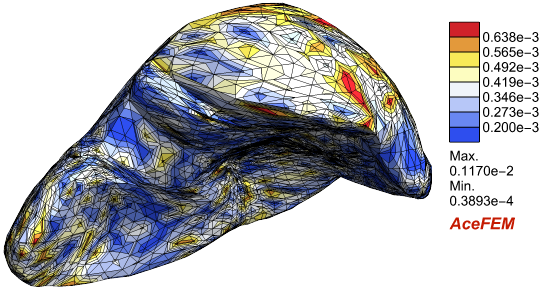}}\hspace{0.04\textwidth}
     
     \subfloat[GP component error, $e_{\text{GP}}$, see Eq.~\eqref{eq: error_gp}]{\includegraphics[width=0.47\textwidth]{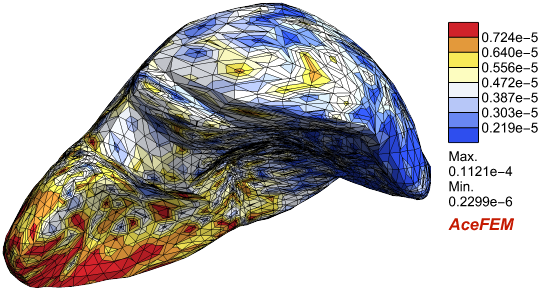}}\hspace{0.04\textwidth}
     \subfloat[Uncertainty prediction, $(u_{\sigma}^{*})^{2}$, see Eq.~\eqref{eq: mean_std_prediction}]{\includegraphics[width=0.47\textwidth]{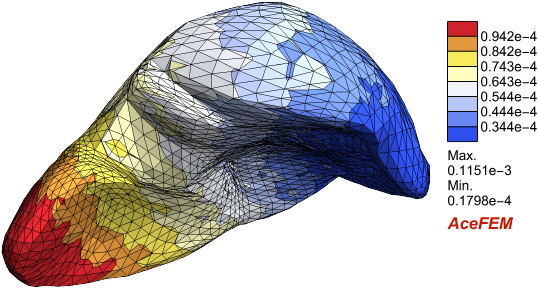}}
    
      \caption{Deformation of the 3D liver for the highest nodal displacement case, subjected to external body force (0.71, 0.69, -0.78) N/kg. Nodal displacements obtained using the (a) proposed framework (b) FEM respectively. (c) Nodal error and (d) reconstruction error of the framework. (e) GP component error. (f) Uncertainty predicted by the framework for the full field solution.}
     \label{fig: liver results}
\end{figure}

In this section we study the performance of the framework. Firstly, we showcase two particular load cases for the 2D and 3D example representing highest nodal displacement examples of each scenario. We visualize the predictions and corresponding errors, which are plotted on deformed meshes. In particular, we utilize the aforementioned breakdown of the framework error into reconstruction- and GP error, which aims to provide a clear and accessible understanding of how each component contributes to the overall performance of the framework. Figures~\ref{fig: 2D beam prediction} and \ref{fig: liver results} illustrate the interpolated node-wise L2 norm of the framework's predictions, including displacements, associated uncertainties, and errors for the 2D beam and 3D liver cases, respectively. Further, we provide overall error metrics for the entire test sets to quantify the framework's performance.

For the 2D beam example, we analyse the test example in with the maximum nodal displacement of 1.16 m, see Figure~\ref{fig: 2D beam prediction}. The prediction of displacement field obtained using the proposed framework which is in a very good agreement with the FEM solution, as presented in the Figures~\ref{fig: 2D beam prediction}a-b. Figure~\ref{fig: 2D beam prediction}c provides nodal error of the entire framework.  While Figure~\ref{fig: 2D beam prediction}d and Figure~\ref{fig: 2D beam prediction}e provide reconstruction error and GP error components as described by Eq.(\ref{eq: decode_error}) and Eq.(\ref{eq: error_gp}) respectively. Error plots show that all the errors are fairly low when compared to the magnitude of the displacement solution thereby displaying the capability of the framework in making accurate predictions. Figure~\ref{fig: 2D beam prediction}f provides nodal uncertainty estimates predicted by the framework (2 standard deviations), as obtained from Eq.(\ref{eq: mean_std_prediction}). One can observe that the GP component errors are within the uncertainty bounds for respective nodal predictions, i.e., in other words, true solution of the framework is contained within the displacement distribution for all the nodes.

Similarly, we analyse the test example of 3D liver case with the maximum nodal displacement of 0.14 m. Figure~\ref{fig: liver results} shows that the framework is able to accurately predict the deformation responses of the 3D liver geometry subjected to external body forces. Figure~\ref{fig: liver results}a, \ref{fig: liver results}b show that the nodal displacements predicted by the framework are in a great agreement with the FEM solution. Figure~\ref{fig: liver results}c, \ref{fig: liver results}d present the framework and reconstruction errors which are extremely low compared to the magnitude of displacements. For the maximum displacement node, the prediction error is $0.2\%$ of its magnitude.  Directing attention to Figure~\ref{fig: liver results}e and Figure~\ref{fig: liver results}f, these figures present the error contribution of the GP component and the predicted uncertainty respectively. Again, the nodal GP errors are within the respective nodal uncertainty values thereby indicating that the framework is able to make reliable predictions.

The performance of the framework over the entire test set is computed using the mean absolute error metric. For the $m^\text{th}$ test case, the mean error is given as follows:  
  \begin{equation}
     e_m = \frac{1}{\mathcal{F}}\sum_{i=1}^{\mathcal{F}}{|\tilde{\bfm{u}}^{*}_{\mu}(\bfm{f}_m)^{i} - \bfm{u}_m^{i}|},
 \end{equation}
where $\mathcal{F}$ is the number of DOFs of the mesh representing the full field space, $\tilde{\bfm{u}}^{*}_{\mu}(\bfm{f}_m)$ is the mean prediction of the GP+Decoder framework (full field prediction) as described through Eq.(\ref{eq: mean_std_prediction}) and $\bfm{u}_m$ is the finite element solution for the full field space. To have a single validation metric over the entire test set, we compute the average mean norm $\Bar{e}$ and the corrected sample standard deviation $\sigma(e)$ as follows: 
\begin{equation}
\begin{split}
     \Bar{e} =\frac{1}{M} \sum_{m=1}^{M} e_m, \qquad
     \sigma(e) = \sqrt{\frac{1}{M-1}\sum_{m=1}^{M} \left(e_m - \Bar{e} \right)^2}.
\end{split}
\label{eq:error_metric_det}
\end{equation}

Finally, in addition to that, we also use the maximum error per degree of freedom over the entire test set
 \begin{equation}
    e_{\text{max}}=\max_{m, i}|\tilde{\bfm{u}}^{*}_{\mu}(\bfm{f}_m)^{i} - \bfm{u}_m^{i}|.
    \label{eq:error_metric_max}
 \end{equation}

The proposed framework predicts probabilistic displacement fields corresponding to the full field mesh. First, we analyse the prediction performance by comparing mean predictions of the framework to the true FEM solutions.  Table~\ref{tab: test_metrics} presents error metrics for both benchmark examples. Additionally, the maximum nodal displacements for both examples are also provided, i.e., we compute displacements of all the nodes for every single test case, and provide the maximum nodal displacement in the table. Table~\ref{tab: test_metrics} shows that both mean and maximum errors are fairly low when compared to the displacement magnitudes, which demonstrates that the proposed framework is capable of predicting mechanical deformation responses with very low errors.

\begin{table}[h]
\begin{center}
 \begin{tabular}{l | c | c | c | c | c  } 
 Example & $M$ & $\Bar{e}$ [m] & $\sigma(e)~[\text{m}]$ & $e_\text{max}$ [m]& Max nodal disp. [m] \\ [0.5ex] 
 \hline
 2D Beam & 300 & 0.1 E-3 & 0.8 E-4 & 7.5 E-3 & 1.16 \\ 
 3D liver & 400 & 0.9 E-4 & 0.3 E-4  & 2.3 E-3 & 0.14 
\end{tabular}
\end{center}
\caption{Error metrics for 2D and 3D test sets for predictions using the proposed GP+autoencoder framework. $M$ stands for the number of test examples, and $\Bar{e}$, $\sigma(e)$ and  $e_\text{max}$ are error metrics defined in Equations~\eqref{eq:error_metric_det}-\eqref{eq:error_metric_max}.}
\label{tab: test_metrics}
\end{table}

\subsection{GP performance in the latent space} \label{sec: model check}

GPs are utilized to predict outcomes in the latent space. The indpendent GPs forecast a Gaussian probability distribution for each latent DOF. To ensure the efficiency and accuracy of GP's predictions, we provide additional accuracy measures. First, we define prediction of a particular DOF as \emph{healthy}, if the corresponding true solution lies within two standard deviations of the mean value. 

\begin{figure}[!ht]
     \centering
     \subfloat[Correct prediction]{\includegraphics[width=0.5\textwidth]{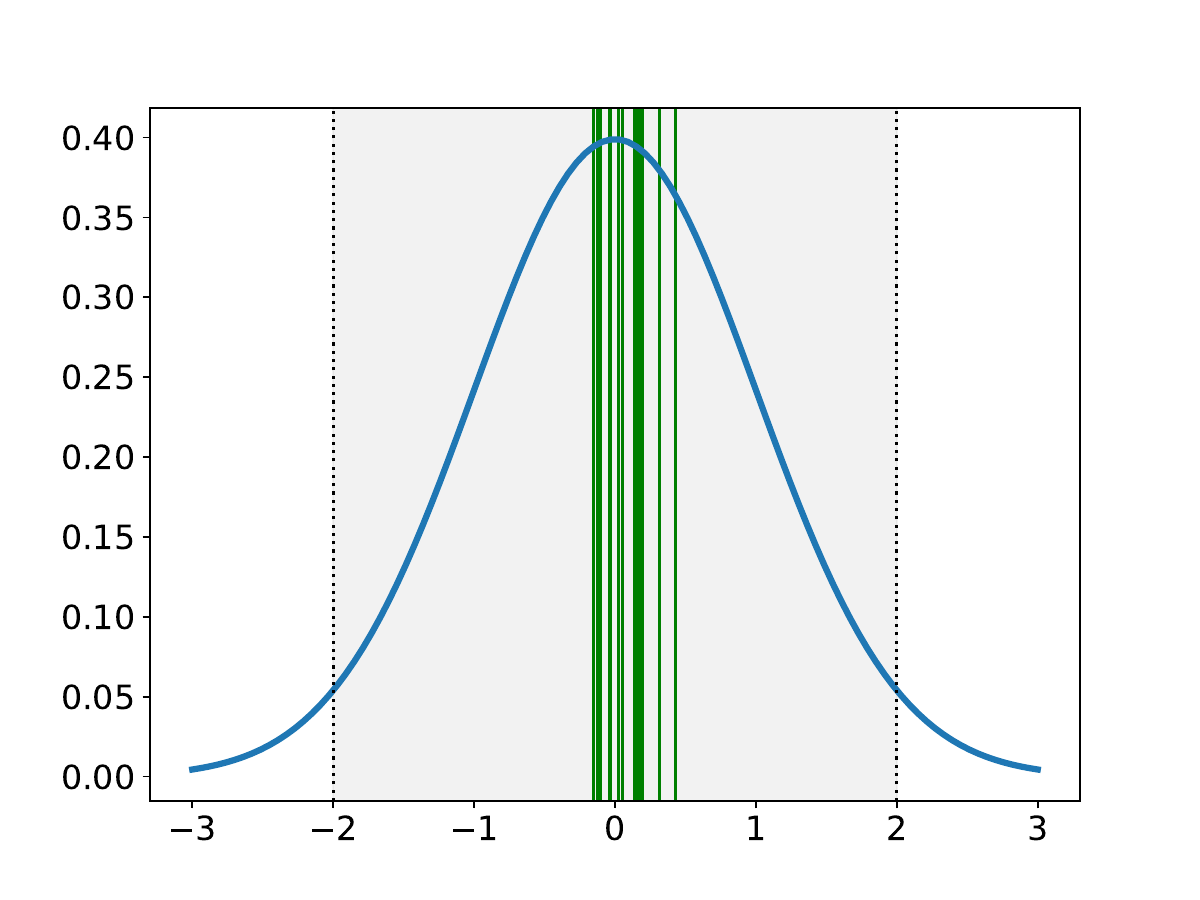}}
     \subfloat[Incorrect prediction ]{\includegraphics[width=0.5\textwidth]{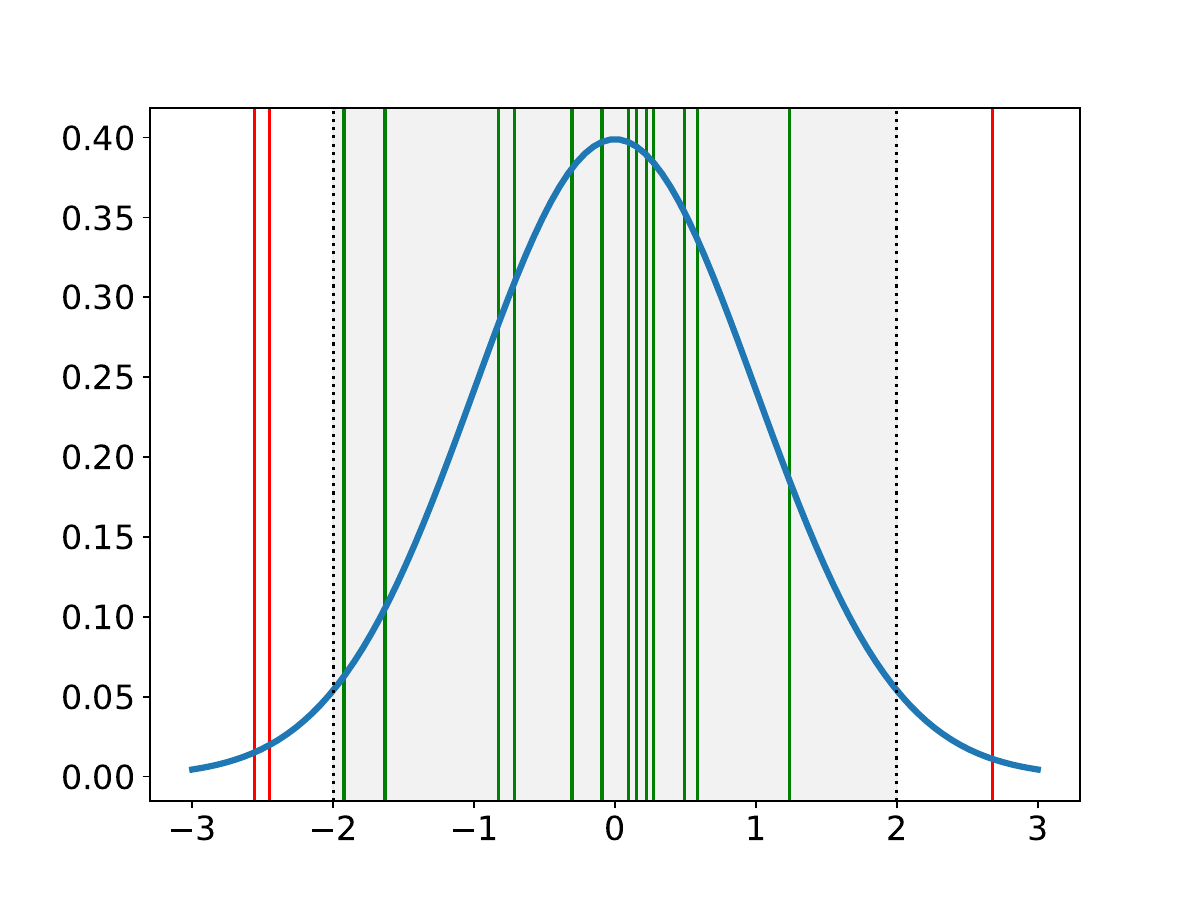}}
      \caption{Latent distribution prediction for two randomly chosen test examples from the liver test dataset. All the 16 true latent values are transformed to an uniform scale, along with the displacement random variables. True latents lying within the two standard deviations ($\pm2\sigma$ region represented by shaded gray region) are indicated with green lines, while the latents lying outside are indicated with red colors. (a) All true latent displacements lie within $\pm2\sigma$, which we classify as a correct prediction.
      (b) Three true latent values don't lie within $\pm2\sigma$, which we classify as an incorrect prediction.}
     \label{fig: correct_prediction}
\end{figure}

To give an example, for the liver case, we plot latent probability distributions for a randomly chosen test example and check if the corresponding true latent values lie within the distribution. These true latent values are derived from the autoencoder compression of full field Finite Element Method (FEM) solutions. The displacement variables for all the latent distributions are transformed to an uniform scale. If the true latent solution for that latent distribution lies within ($u_\mu \pm 2u_\sigma$), it is indicated by the green line, otherwise with the red line as shown in the Figure~\ref{fig: correct_prediction}. For a particular test examples, if all the latent predictions are \emph{healthy}, then that prediction is regarded as \emph{correct}, see, e.g., Figure~\ref{fig: correct_prediction}a. Even if a single latent prediction is not \emph{healthy}, that prediction is regarded as \emph{incorrect}, see Figure~\ref{fig: correct_prediction}b. This is because a single non-healthy latent prediction can affect the full field solution during the reconstruction step.

The percentage of \emph{healthy} and correct predictions together contribute to a comprehensive assessment of the GP's predictive capabilities in predicting latent distributions. Table~\ref{tab: gp_metrics} provides these accuracy measures for both training and test datasets of both cases. The study on test datasets reveals that even though $\sim 95\%$ of predicted latent components are \emph{healthy}, the percentage of \emph{correct} GPs predictions can be significantly lower. We expect this unwanted effect to be more pronounced with the increasing dimension of the latent space.

\begin{table}[!h]
\begin{center}
\renewcommand{\arraystretch}{1.2}
 \begin{tabular}{l | c | c | c | c |  } 
 \multirow{2}{*}{Case} & \multicolumn{2}{c|}{\% correct predictions} & \multicolumn{2}{c|}{\% healthy predictions}\\ \cline{2-5} 
      &  Train & Test & Train & Test \\
\hline
 2D beam & 95.1 & 77.7  & 99.1 & 94.1\\ 
 3D liver & 93.1  & 74.5 & 99.0 & 95.6
\end{tabular}
\end{center}
\caption{Performance of the GP in correctly predicting latent distributions of both cases.}
\label{tab: gp_metrics}
\end{table}

\subsection{Uncertainty for missing data region}
\label{sec: missing data region}

\begin{figure}
     \centering
     \subfloat[]{\includegraphics[width=0.35\textwidth]{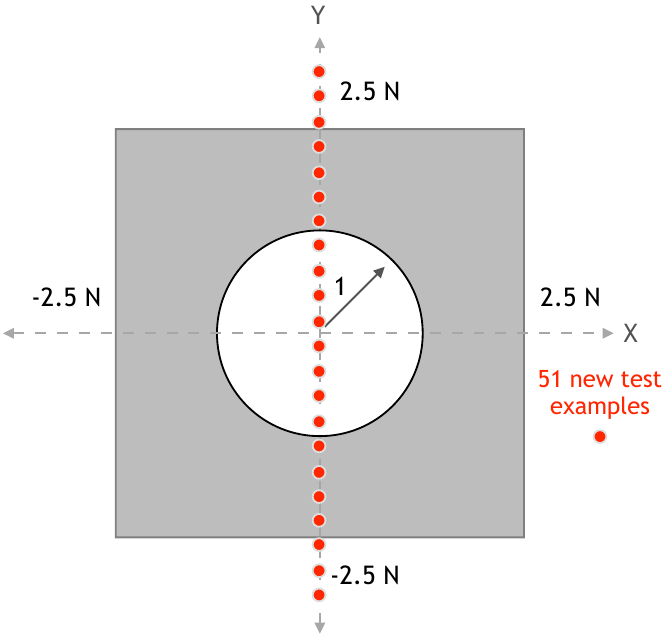}\label{fig: extrapolated a}}\hspace{0.15\textwidth}
     \subfloat[]{\includegraphics[width=0.35\textwidth]{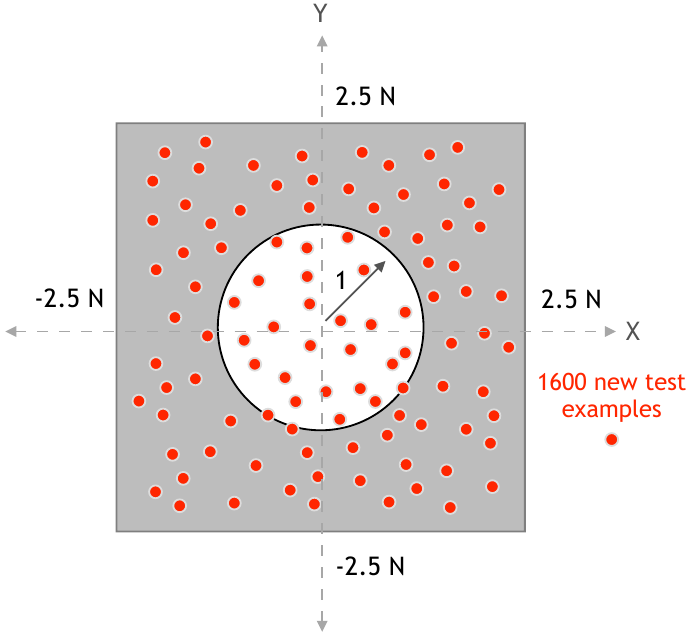}\label{fig: extrapolated b}}
      \caption{Reduced training dataset. The subset of the original dataset is marked with gray color. The distributions of two new test sets are illustrated with red dots in (a) and (b).}
     \label{fig: extrapolated}
\end{figure}

As presented in the Table~\ref{tab:datasets}, the original 2D beam dataset of 6000 examples was generated by applying random forces of magnitude -2.5 N to 2.5 N in X and Y-direction. In order to further test the capability of the framework, we remove all the cases whose nodal input force magnitude is less than 1~N (the region represented by white circle in Figure~\ref{fig: extrapolated}). This subset of original data (5019 examples) is now used to train the proposed framework. Now the framework trained on this masked dataset is used to make predictions on two new input datasets. 

\begin{figure}[!h]
     \centering
    \subfloat[]{\includegraphics[width=0.42\textwidth]{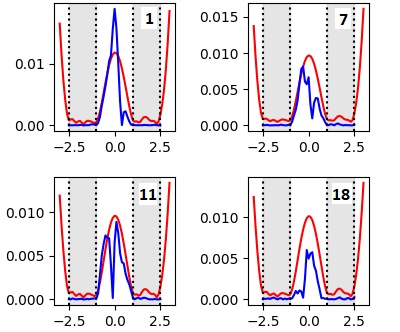}
    \label{fig: latent_uncertainty_1D_4dofs}}
    \hspace{0.09\textwidth}
    \subfloat[]{
    \includegraphics[width=0.42\textwidth]{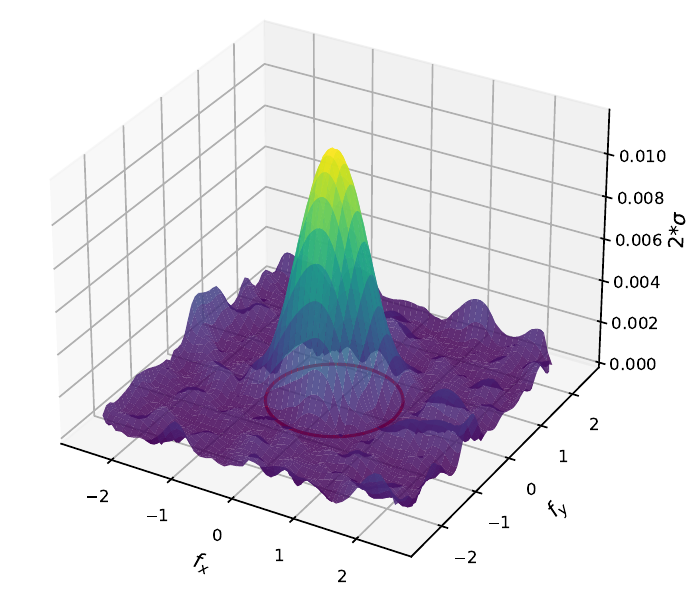}
    \label{fig: latent_uncertainty_3D_0dof}
    }
     \caption{Uncertainty prediction for 2D beam case, for the GP trained using the missing dataset as indicated in Figure~\ref{fig: extrapolated}. In (a), the uncertainties (red line) and true latent prediction errors (blue line) are shown for selected four latent components (component number is indicated in the inset) for the test data depicted in Figure~\ref{fig: extrapolated a}. In (b), the uncertainties are shown for the first latent component for the test data depicted in Figure~\ref{fig: extrapolated b}. In (a) and (b) we observe the trend of increased uncertainty in the region not supported by the training data. }
\end{figure}

In the first case, we apply a one dimensional Y-directional force whose magnitude changes from -3 to 3 N, as indicated by red dot line in the Figure~\ref{fig: extrapolated a}. In Figure~\ref{fig: latent_uncertainty_1D_4dofs}, we plot the uncertainty of selected latent DOF predicted by GP. For comparison, we also plot the error of latent DOF prediction as well. The figure shows that the uncertainty of prediction for all the latent DOFs is low in the region supported by the data and the prediction errors are within uncertainty bounds. The uncertainty grows quickly in both interpolated and extrapolated missing data region. Especially, the uncertainty grows rapidly in the extrapolated region (for magnitude of the force greater than 2.5 N). 

In the second case, we use a new test dataset in which two dimensional input forces are randomly generated as denoted by red dots in Figure~\ref{fig: extrapolated b}. In Figure~\ref{fig: latent_uncertainty_3D_0dof} we plot the uncertainty for the first latent DOF. It is evident that the uncertainty shows an increasing trend in the region not supported by training data. All the 20 latent DOFs show a very similar trend. In the presented examples, these uncertainties correlate with the GP prediction errors, which can provide useful information. However, it's crucial to emphasize that the predictive capabilities in the data-sparse and extrapolated regions can never be assured, with the tendency to deteriorate with the distance from the training data. 

\begin{figure}[t!]
     \centering
     
     \subfloat[Framework solution]{\includegraphics[width=0.47\textwidth]{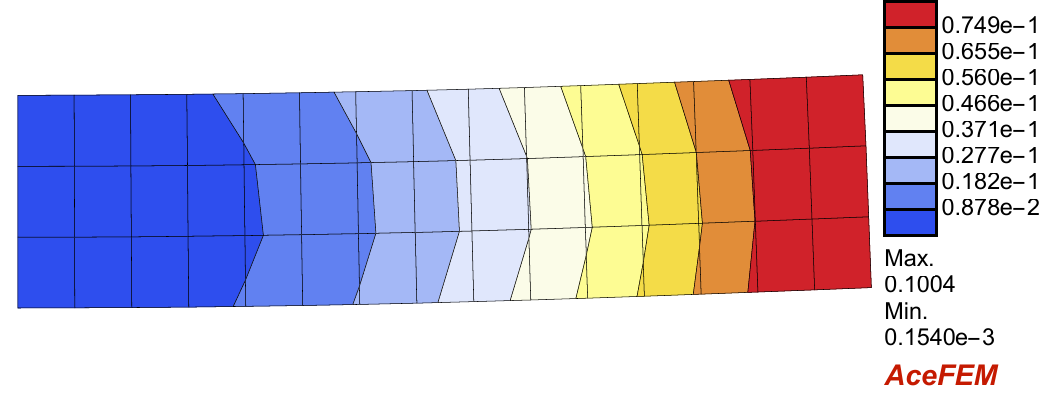}}\hspace{0.04\textwidth}
     \subfloat[FEM solution]{\includegraphics[width=0.465\textwidth]{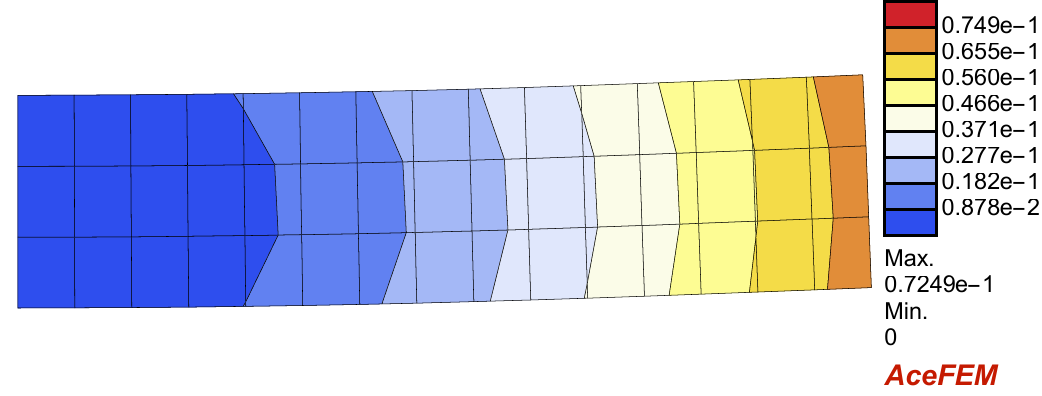}}
     
     \subfloat[Absolute error, $e_\text{f}$, see Eq.~\eqref{eq: absolute error}]{\includegraphics[width=0.47\textwidth]{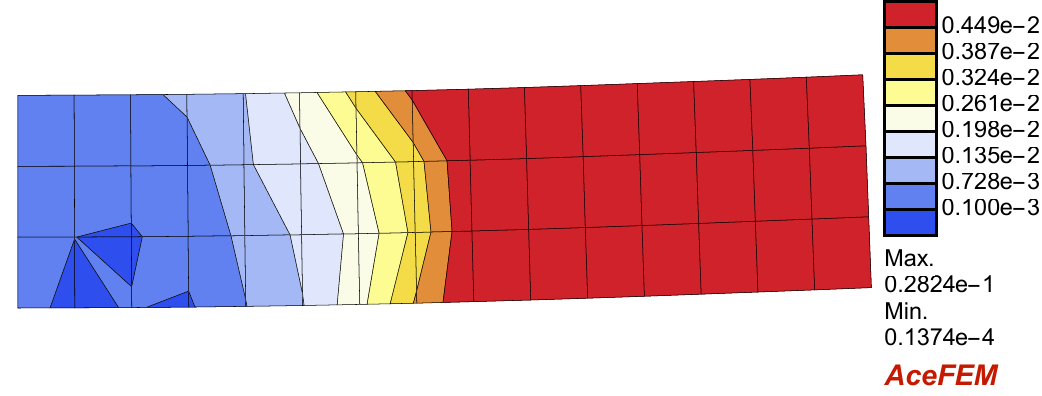}}\hspace{0.04\textwidth}
    \subfloat[Reconstruction error, $e_r$, see Eq.~\eqref{eq: decode_error}]{\includegraphics[width=0.47\textwidth]{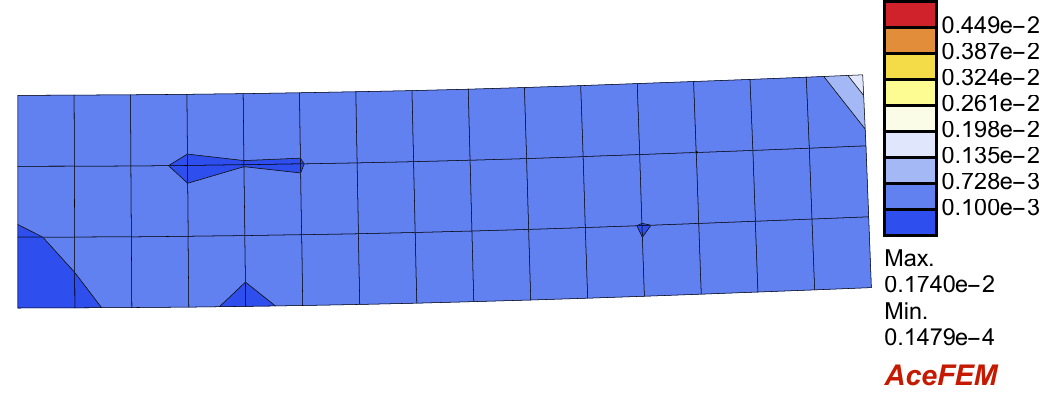}}\hspace{0.04\textwidth}
     
     \subfloat[GP component error, $e_{\text{GP}}$, see Eq.~\eqref{eq: error_gp}]{\includegraphics[width=0.47\textwidth]{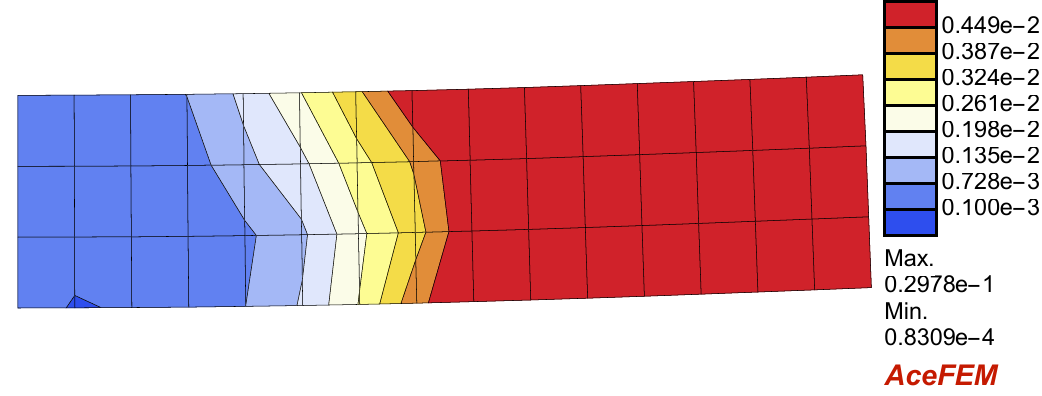}}\hspace{0.04\textwidth}
     \subfloat[Uncertainty prediction, $(u_{\sigma}^{*})^{2}$, see Eq.~\eqref{eq: mean_std_prediction}]{\includegraphics[width=0.47\textwidth]{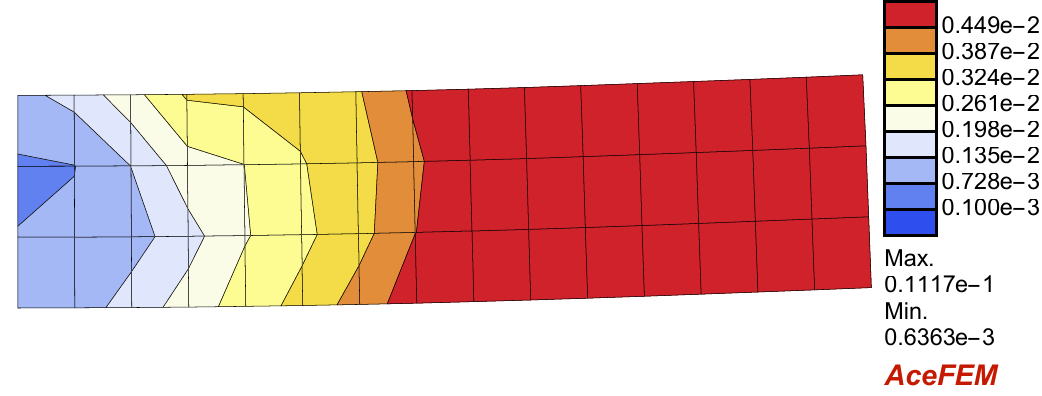}}
    
      \caption{Deformation of the 2D beam subjected to a point load (-0.14, 0.15) N on the corner node, which is present in the missing training data region as described by white circle in Figure~\ref{fig: extrapolated b}. Nodal displacements obtained using the (a) proposed framework (b) FEM respectively. (c) Nodal error and (d) reconstruction error of the framework. (e) GP component error. (f) Uncertainty predicted by the framework for the full field solution. }
     \label{fig: 2D missing full prediction}
\end{figure}

Similarly, higher uncertainty values in regions not supported by data are also observed in the full field space. For instance, one of the test points located in the white region of Figure~\ref{fig: extrapolated b} has its prediction projected into the full space. As shown in Figure~\ref{fig: latent_uncertainty_3D_0dof}, the latent uncertainty for such an input is high. Similarly, Figure~\ref{fig: 2D missing full prediction}f shows that the predicted full field uncertainties are also high throughout the 2D domain. As mentioned earlier, uncertainties are only responsible for tracking the GP component error and are not directly correlated with the autoencoder errors. However, higher levels of uncertainty values, which exceed acceptable engineering thresholds, serve as a reliable indicator of whether the full field solution should be trusted or not.

\section{Conclusions}
\label{sec: conclusion}
This work presents a novel approach to probabilistic surrogate modeling of high-fidelity simulations. It combines Gaussian processes with reduced-order modeling to efficiently simulate the mechanics of solids while accounting for uncertainties. The reduced representations of high-dimensional displacement data are achieved using autoencoder neural networks, allowing for drastic non-linear compression. For instance, this method reduced approximately 9100 degrees of freedom (DOFs) to 16 latent-space components, as demonstrated with the liver example. This strategy significantly reduces the computational cost of probabilistic predictions using Gaussian processes by lowering the output dimensionality, which is crucial because this cost can scale linearly to cubically with the output dimension size. We demonstrate that the proposed framework can accurately predict non-linear hyper-elastic deformations of solid bodies, along with the associated model uncertainties. 

In addition to demonstrating the capabilities of our probabilistic framework, we also studied its possible limitations. We emphasized that while the uncertainties in the missing data region provide a useful indicator that is correlated with the distance from training data, they do not necessarily need to represent the true solution errors. The development of methods to improve the quantitative nature of those predictions remains an open topic for future research. We also pointed out that even if a single component of the latent solution is predicted incorrectly, it can impact the entire reconstructed solution. This led us to propose more meaningful error metrics for quantifying the performance of the proposed probabilistic framework. This issue can be alleviated by developing a suited training strategy, which poses an interesting future research direction.  

This study opens up several other opportunities for future research. One exciting direction is applying this approach to probabilistic simulations of time and path dependent problems. One could use latent state to track how solutions evolve in such cases, which would make the whole process much faster and more efficient \citep{Nikolopoulos2022}. Another direction would be to exploit state of the art graph autoencoder networks to find compressed representations for arbitrary high-dimensional meshes \citep{barwey2023multiscale}. This advancement would overcome the limitations of fully connected networks, which are recognized for their challenges in handling problems with a high number of dimensions. Finally, we believe that the proposed approach can find direct application in research and development, which is not restricted to problems in mechanics but covers a broader spectrum of engineering and scientific domains. To facilitate this, we have made all the codes and datasets used in this work available as open-access in the GitHub repository at \href{https://github.com/saurabhdeshpande93/gp-auto-regression}{https://github.com/saurabhdeshpande93/gp-auto-regression}.

\section*{Contributions}
\textbf{Saurabh Deshpande}: Conceptualization, Data curation, Formal analysis, Methodology, GP+ANN implementation, Validation, Visualization, Writing – original draft, Writing – review \& editing. \\
\textbf{Hussein Rappel}: Formal analysis, Methodology, GP implementation, Writing – original draft, Writing – review \& editing. \\
\textbf{Mark Hobbs}: Reviewing the manuscript.\\ 
\textbf{Stéphane P.A. Bordas}: Funding acquisition, Reviewing the manuscript, Supervision.\\  
\textbf{Jakub Lengiewicz}: Conceptualization, Data curation, Investigation, Methodology, Project administration, Supervision, Validation, Visualization, Writing – original draft, Writing – review \& editing.

\vspace{5mm}

\emph{Acknowledgements:}
\begin{wrapfigure}{l}{0.34\textwidth}
    \includegraphics[width=0.3\textwidth]{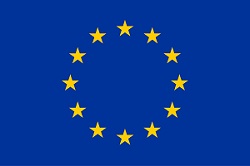}
\end{wrapfigure}
This project has received funding from the European Union’s Horizon 2020 research and innovation programme under the Marie Sklodowska-Curie grant agreement No. 764644. 
Jakub Lengiewicz would like to acknowledge the support from EU Horizon 2020 Marie Sklodowska Curie Individual Fellowship \emph{MOrPhEM} under Grant 800150. Jakub Lengiewicz would like to acknowledge the support of the Luxembourg National Research Fund (FNR) through the project SUMO, Grant INTER/GACR/21/16555380. St\'ephane Bordas and Jakub Lengiewicz are grateful for the support of the Fonds National de la Recherche Luxembourg FNR grant QuaC C20/MS/14782078. St\'ephane Bordas received funding from the European Union's Horizon 2020 research and innovation programme under grant agreement No 811099 TWINNING Project DRIVEN for the University of Luxembourg. This paper only contains the author's views and the Research Executive Agency and the Commission are not responsible for any use that may be made of the information it contains.

\bibliography{mybibfile}
\end{document}